\documentclass[a4paper,fleqn]{cas-dc}

\usepackage[authoryear]{natbib}
\usepackage{xcolor}
\newcommand{\hide}[1]{}

\def\tsc#1{\csdef{#1}{\textsc{\lowercase{#1}}\xspace}}
\tsc{WGM}
\tsc{QE}
\tsc{EP}
\tsc{PMS}
\tsc{BEC}
\tsc{DE}

\begin{document}
\let\WriteBookmarks\relax
\def\floatpagepagefraction{1}
\def\textpagefraction{.001}
\shorttitle{V-shape family detection efficiency}
\shortauthors{Deienno et~al.}


\title [mode = title]{Efficiency characterization of the V-shape asteroid family detection method}      




\author[1]{Rogerio Deienno}[type=editor,
                        orcid=0000-0001-6730-7857]
\cormark[1]
\ead{rdeienno@boulder.swri.edu}
\ead[url]{https://www.boulder.swri.edu/\~rdeienno}


\address[1]{Department of Space Studies, Southwest Research Institute, 1050 Walnut St., Boulder, CO 80302, USA}

\author[1]{Kevin J. Walsh}[style=, orcid=0000-0002-0906-1761]
\ead[url]{https://www.boulder.swri.edu/\~kwalsh}

\author[2]{Marco Delbo}[%
   role=,
   suffix=,
   ]
\ead[URL]{https://www.oca.eu/en/marco-delbo}


\address[2]{Observatoire de la C\^ote d'Azur, CNRS-Lagrange, Universit\'e C\^ote d'Azur, CS 34229, 06304 Nice Cedex 4, France}





\begin{abstract}
Following the break up of a parent body, the Yarkovsky effect causes asteroid family members to spread in orbital semimajor axis with a rate often inversely proportional to their diameter. This size dependent semimajor axis drift causes family members to form structures in the semimajor axis vs inverse diameter plane that have the shape of the letter V. The V-shape method consists in finding the borders of such V-shapes of unknown center and opening. 
Although successfully employed to find some very old families in the inner main asteroid belt, the V-shape searching method is very sensitive to many parameters.
In this work, we first created and evolved a synthetic asteroid family over billions of years. Then, by adding uncertainties to semimajor axis and diameter of the evolved synthetic family components, we randomly generated additional 99 similar, but not perfectly V-shaped, family clones.  We chose a fairly low initial velocity dispersion of 20 m/s for our family. Thus, we can more easily relate the spreading in semimajor axis with the family's age (slope of the evolving V). A synthetic background with an initially randomly distributed components was also created and evolved for 100 Myr. 
Thus, by setting different levels of ratio of the synthetic family and background asteroids, we derived a detection efficiency map for the V-shape method and determined how sensitive the results can be based on signal-to-noise levels. 
We also determined optimal parameter values for the method's efficiency. We found that, families older than $\approx$3 Gyr are likely undetectable, with a method efficiency of 50\% or less, whereas younger families (0.5--2.5 Gyr)  are more easily detected by the method, with an efficiency of $\gtrsim$80\%.




\end{abstract}




\maketitle

\section{Introduction}

The asteroid main belt is one of the best tools to study the entire history of our solar system. 
It traces migration of the giant planets \citep{levison2009,walsh2011,morbidelli2015,vokrouhlicky2016}, contains sample of nearly the entire suite of primordial solar system material \citep{demeo2015,johansen2015} and records epochs of major upheaval and collisions \citep{bottke2015}.
These collisions created families of fragment asteroids which can be still identified and dated. 
These collisional families can be used to trace the evolution of the main belt throughout its history as well as epochs of asteroid bombardment on other asteroids and the terrestrial planets \citep[e.g][]{bottke2015,vokrouhlicky2017}.

However, the completeness of knowledge of families in the asteroid belt is currently unknown \citep{nesvorny2015}. The census of known families is clearly incomplete as a function of time, across all types of asteroids, and across all regions of the asteroid belt. While existing techniques are adequate for analyzing/characterising known families, they have clear limitations on finding small or old families \citep{broz2013,nesvorny2015}.

 After their formation in catatstrophic collision \citep[see][]{michel2015} though the collisional remnants' orbital elements remain similar, they evolve and spread over time \citep{nesvorny2015}. Tightly grouped clusters of asteroids have been recognized as families for over 100 years \citep[see also \citealp{gradie1979,valsecchi1989}]{hirayama1918}. These families of asteroids are detected by their similar orbital elements, and the number of candidate families has increased with the increase in cataloged asteroids. \citet{zappala1995} found 30 families in a sample of $\sim$12,000 asteroids, \citet{nesvorny2015} found $\sim$122 families in a sample of 350,000, \citet{masiero2013} found 60 using over 110,000 and \citet{milani2014} found over 100 candidate families from 300,000 known asteroids (just to name a few of the many works on this topic).

The ages and sizes of the known asteroid families in current catalogs have a curious distribution. There are some very old families with very large parent bodies (with ages older than Gyr; Themis, for example) and numerous young families (with ages younger than Gyr) with a wide range of parent body sizes \citep[see][or \citealp{spoto2015}]{broz2013}. However, there is a significant lack of families around large asteroids (D > 100 km) with ages between 1 and 2.5 Gyr. For ages older than 1.5 Gyr old, there are none known around smaller parent bodies (D < 100 km). However, smaller and older families are hard to detect due to collisional evolution of the family and spreading of their orbits. Therefore, it is not clear what aspects of the asteroid family record over time is real and what is simply an artifact of our detection techniques and tools.

Most current efforts have been focused on detecting and characterizing families through any means possible - at times folding in astronomical data such as broadband visible colors from SDSS or albedos from WISE \citep{parker2008,masiero2013}. Here, we aim to characterize and calibrate one detection technique in order to, in a future work, be able to de-bias the families that it detects (the latter will be a future step and is, therefore, out of the scope of the present work). This will likely not find as many families as presented in some of the previously mentioned databases, but it will inform on the efficiency (likelihood) of their detection and thus provide an idea of how many families are still  undetected, thus what the entire population could look like.

\subsubsection*{Hierarchical Clustering Method}

One standard tool used to identify and analyze asteroid families has been the Hierarchical Clustering Method (HCM: see reviews by \citet{bendjoya2002,nesvorny2015}). This technique starts with a possible parent asteroid and tests to see if any asteroid's orbit is within a critical ``distance'', where the distance is measured in the velocity ($V_C$) required to move a body between the two orbits (using proper orbital elements, $a,e,i$). If a neighboring asteroid is within the limiting distance, then the same check is run from this neighbor and the family expands. Starting with a proposed parent main belt asteroid as the center of the family, the linking criteria $V_C$ is increased and the number of linked bodies is counted at each increment. 

This method has been widely used, and in some cases the linking metric has been expanded to include asteroid physical properties, such as albedo \citep{masiero2013} or their photometric colors \citep{parker2008,carruba2013}. The basic method, and heart of nearly all previous work, relies exclusively on the three primary orbital elements of semimajor axis ($a$), eccentricity ($e$) and inclination ($i$). However, asteroid orbits spread in semimajor axis over time due to the thermal forces of the Yarkovsky effect. This effect is strongly size-dependent ($\tau_{yark} \sim$ 1/D), so the more numerous smaller family members disperse faster than the fewer large bodies (see Fig. \ref{fig2}). On 100 Myr timescales $\sim$1 km bodies can drift $\sim$0.01 au, and can drift $\sim$0.1 au in $\sim$1 Gyr.

Meanwhile, the smaller objects drifting the fastest are also subject to collisional evolution at the hands of the rest of the asteroid belt. A 1 km object has a collisional lifetime of only $\sim$500 Myr, and a 0.1 km object a lifetime of only $\sim$50 Myr \citep{bottke2005a,bottke2005b}. Thus over time the family spreads out and its smaller members are possibly disrupted, which then should produce numerous new smaller fragments, but with a range of sizes, shapes and obliquity.

Similarly, as objects drift away from the center of the family they can experience diffusion of eccentricity and inclination due to interactions with minor orbital resonances throughout the asteroid belt \citep{vokrouhlicky2017}. The magnitude of this effect, and how it changes the efficacy of family detection algorithms will depend on where in the asteroid belt the family has formed and the array of resonances nearby.

Young families have always been easier to detect using HCM because they are more tightly clustered owing to less Yarkovsky orbital drift. The surveys of the asteroid belt using HCM show the clear abundance of small young families relative to big old families \citep[Fig. 4]{broz2013}. Furthermore, it is clear that HCM does not capture all of the small asteroids that are members of the family. As found in \citet{parker2008} when selected members are removed, there are still very clear ``halos'' of similarly-colored bodies surround the core of the detected family. 

\subsubsection*{V-shape detection}

\begin{figure*}
 	\centering
 		\includegraphics[scale=0.35]{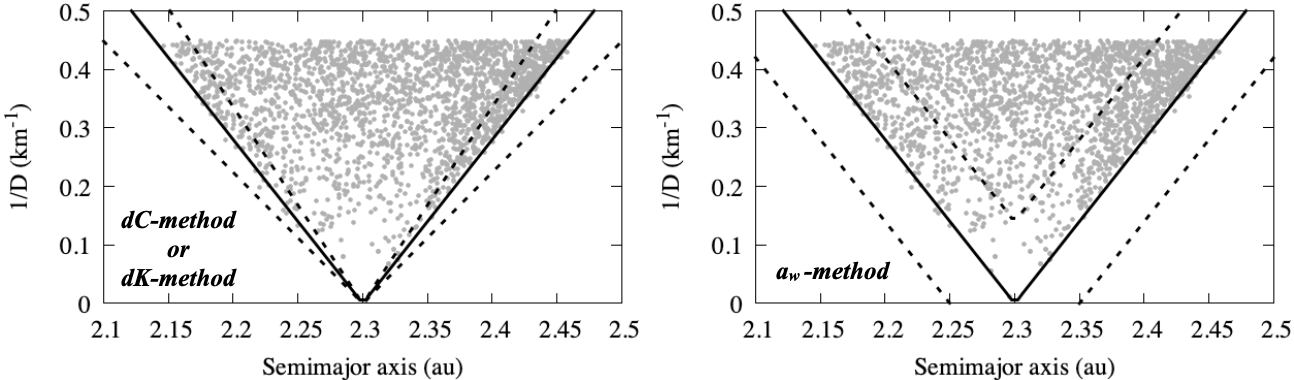}
 	\caption{Representation of a perfectly V-shaped synthetic family in the semimajor axis (au; x-axis) vs 1/D (km$^{-1}$; y-axis) plane. Gray dots represent the members of a fictitious family. Solid black line denotes the nominal expected V shape for this family. Left: representation of the $dC$-$method$ or $dK$-$method$ with an applied $dC$; $K\pm \Delta K = \ \sqrt{p_v}/[D_0(C\pm dC)]$ (dashed lines). Right: representation of the $a_{\rm w}$-$method$ with a shift of $\pm a_{\rm w}$ in $1/D$ (dashed lines).}
 	\label{fig1}
 \end{figure*}

The evolution of asteroid's semimajor axis as a function of their size creates the identifiable ``V'' shape (see Figure \ref{fig1}) when plotted in $H~vs~a$ or $1/D~vs~a$. This effect is simply showing the 1/D dependence in asteroid drift rates (see \citet{bolin2018b} for variations on the simplistic 1/D dependence based on thermal inertia effects). \citet{walsh2013} used the formulations of \citet{vokrouhlicky2006} where the distance in semimajor axis of two asteroids was normalized by their size to determine the Yarkovsky drift distance of all neighboring asteroids. Re-scaling asteroid's distance from each other in a size-dependent way is typically referred to as the $C$ parameter \citep{vokrouhlicky2006}. The $C$ distribution of asteroids was used by \citet{walsh2013} to locate the center of the family, based on a cluster of similar $C$ values for a given tested family center location, similar to what \citet{vokrouhlicky2006} did for numerous families. A bounding $C$ value for the family can then be used to estimate the upper limit age for the family, based on the maximum Yarkovsky drift timescales.

As found in numerous works the distribution in $C$ is not a perfect spike or peak, rather its a gaussian-like distribution that stands above an otherwise flat background \citep[see][Fig. 1]{vokrouhlicky2006}. This has opened up numerous approaches to detection of families' V-shapes amongst the background of unassociated asteroids, and V-shape finding algorithms have been formulated with (at least) two variations \citep{bolin2017}. The first is the "border method" that utilizes a ratio between the number of objects inside and outside of a V-shape drawn in $a~vs~H$ or $a~vs~1/D$ space. This is sensitive to detecting families that have a strong or distinct edge and has good results against a small background of asteroids \citep[see][or \citealp{delbo2017} for examples]{walsh2013}. The second technique explored is the "density method" that aims to detect a peak of asteroid density along a V-shape, rather than detecting its edge (the peak of the gaussian rather than its edge). \citet{bolin2017} found that this had better results for families embedded in a heavy background of asteroids.

The border technique has been implemented in a few different ways. The first; called the $dC$-$method$ or $dK$-$method$ herein; is to test a family center and age or slope (where slope is how open the V-shape is, which correlates directly to its age and time spent drifting apart by the Yarkovsky effect). This essentially tests a set of family center ($a_c$) and family slopes $K$, by evaluating a ratio between those bodies just on the inside and those just on the outside of the $V(a_c,C)$ defined V-shape. The slope $K$ relates to the parameter $C$ as $K=\sqrt{p_v}/(D_0C)$, where $p_v$ is the geometric visible albedo and $D_0 =$ 1329 km  \citep{walsh2013,delbo2017}. The distance on either side of the $V$ to use for the test is set by $dC$; how wide of a sliver on either side do we use to count bodies \citep[see][and \citealp{bolin2017}]{walsh2013}. 

A similar but slightly different approach is to offset the V-shape vertically in $a~vs~1/D,H$ space and calculate the ratio of those asteroids above and below the tested $V(a_c,C)$ shape. The offset is established as a delta in $1/D$, and named as $a_{\rm w}$ by \citet{delbo2017}, and this is called the $a_{\rm w}$-$method$ in this work \citep[see][]{delbo2017}. In other words, the $dC/dK$-$method$ search for $(K\pm \Delta K)|a-a_c|$ and the $a_{\rm w}$-$method$ search for $K(|(a-a_c)| \pm a_{\rm w})$ with $1/D, \Delta K, a_{\rm w}$ $>$ 0 (figure \ref{fig1}).

As can be seen in figure \ref{fig1}, although similar, both methods present a major difference in the number of family objects counted above and below the nominal V (black solid line). Therefore, it is also expected that the detection of a certain family will be influenced by this difference.

Here, the goal is to quantify the efficiency of the V-shape detection techniques. In the {\it Methods Section} we describe the forms of the V-shape technique used in the tests and the data sets used for testing.

\section{Methods}\label{methods}

\subsection{Detection Tools}\label{tools}

Both versions of the V-shape technique are more complicated to implement than described in the Introduction (see \citealp{bolin2017} for an exhaustive description of the implementation of the $dC$-$method$ technique, and \citealp{delbo2017} for the $a_{\rm w}$-$method$). To detect the edge of a possible family we are interested in the boundary of the $C$ distribution, or the envelope of Yarkovsky lines for $H$ as a function of $a$. \citet{walsh2013} employed a fitting routine whereby the value of $C$ was varied and the ratio of asteroids with $C-8\times 10^{-6}$ au was compared with $C+8\times 10^{-6}$ au. A strong contrast in numbers indicates the boundary of the family has been reached. Due to the increasing number of asteroids at greater $H$ (smaller sizes), \citet{walsh2013} measured this ratio for three different size ranges, 13.5 < $H_i$ < 15 < $H_{ii}$ < 16 < $H_{iii}$ < 16.5. This simplistic approach has been adapted in different ways since, but highlights some of the configurable parameters: selecting $dC$, how to deal with the Size Frequency Distribution (SFD) of the asteroid population and how to score a fit.

\citet{walsh2013} selected $dC$ in an ad hoc manner, but since then work by \citet{bolin2017} show that results can be quite sensitive to this value, and that different values are more appropriate for different age families. This is actually intuitive, where as a family gets younger, its $C$ decreases and a static $dC$ will become a larger fraction of the tested value $C$. Older families also have more time to spread out where other thermal effects \citep[YORP-induced obliquity variations, see][]{vokrouhlicky2006} can decrease the contrast of the family edge against the background. In this work we explicitly test this and try to define which values are optimal as a function of $C$.

\citet{bolin2018a} took a more elegant approach whereby asteroids in each sliver where weighted by the asteroid belt SFD, such that the few number of larger asteroids were 
weighted more than the many more smaller asteroids. Which SFD is to be used; that presumed for a family just after formation, or the current SFD of the asteroid belt; is one parameter, where minimum and maximum sizes of asteroids are potentially additional parameters.

For each approach there are values associated with the asteroids inside and outside of the V-shape, which may or not be weighted (we do not consider weighting in this work). A simple ratio of these numbers is dangerous as depending on where in the asteroid belt and the size of $dC/dK/a_{\rm w}$ there could be zero bodies in one bin leading to a division by zero (also known as {\it edge effect}). For the specific task of locating and characterizing a very ancient family against a very small background \citet{delbo2017} utilized a $N_{in}^2/N_{out}$ scoring, whereas \citet{bolin2017}
~primarily
~employed 
~$N_{in}/N_{out}$ scoring. Due to the fact that $N_{in}^2/N_{out}$ scoring provided a clear detection of a very old family in \citet{delbo2017}, following works adopted such metric \citep{bolin2018a,delbo2019}. 

Another complicating factor regarding the implementation of the V-shape technique that influences its detection translates as the potential asymmetric evolution of the family. As family members drift due to the Yarkovsky effect expanding them away from the family center, they may encounter mean motion resonances (MMR) with some of the solar system planets, especially Jupiter and Mars. Such MMR encounters can not only easily disperse those objects in the $a,e,i$ space but also cause their ejection from the asteroid belt entirely. With its members getting dispersed and/or lost over time, an asteroid family initially showing a clearly shaped V may also either become unrecognized or partially recognized. The strong resonant interactions on only one side of the V-shape can also make the families to become asymmetric or entirely one-sided  (see Fig. 1 in \citealp{vokrouhlicky2017} for the case of the Flora family and Fig. 1 in \citealp{delbo2017} for the cases of Eulalia, Polana and Primordial families).

\begin{figure*}
 	\centering
 		\includegraphics[scale=0.35]{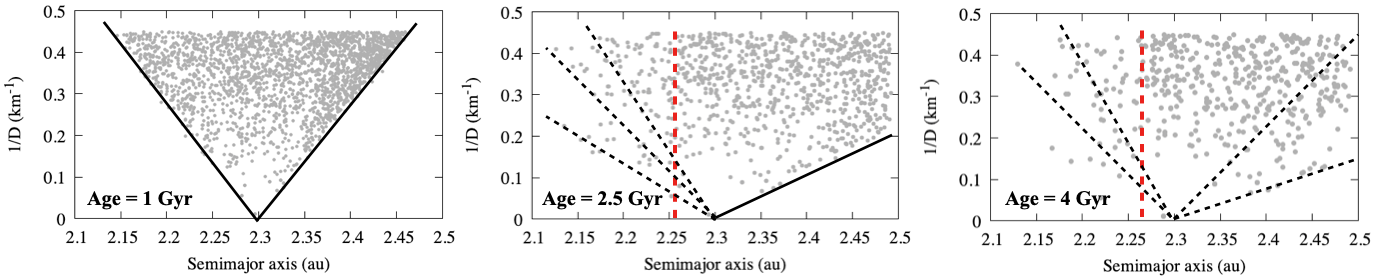}
 	\caption{Dynamical evolution of a synthetic family. Gray dots represent family members. Left: same family shown in figure \ref{fig1} (age equal 1 Gyr) with a representation of the expected nominal V (solid black line). Center: same family on the left for an age equal 2.5 Gyr. Right: what is left from the family after 4 Gyr of evolution. Red: location of the 7:2 MMR with Jupiter, also equal to 9:5 MMR with Mars. Dashed black lines: possible but highly uncertain fits for the nominal V of this family.}
 	\label{fig2}
 \end{figure*}

In figure \ref{fig2} we show an example of the evolution and possible V-shape fit for one synthetic family evolved for 4 Gyr (see section \ref{sfam} for details on synthetic family setup and Yarkovsky drift assumptions). This synthetic family was evolved under the influence of all solar system planets from Venus to Neptune, assumed to be in their current orbits from the beginning of the simulation. 

As can be seen in figure \ref{fig2}, for the 1 Gyr old synthetic family considered the nominal V is very well defined and visible. After 2.5 Gyr, however, although the right side of the family can still be recognized, its left side gets very fuzzy mainly as a consequence of the crossing with the 7:2/9:5 MMR with Jupiter and Mars respectively (note that, in addition to the 7:2/9:5 MMR this region is also affected by important small (high order) weak MMRs, see in particular figure 9 of \citealp{morbidelli1999}). Although these interactions are not strong enough to deplete and erode the family members, the complex resonant structure around $\sim$2.25 au is capable of generating a very large dispersion and some significant loss of bodies. The dashed black lines simply show that this dispersion has made a simple by-eye fit much more challenging. Therefore, in such a situation scoring each side independently could have some advantages. On the other hand, by considering only the right side, and scoring only the objects from a search within  $(K\pm \Delta K)(a-a_c)$ or $K(a-a_c \pm a_{\rm w})$ with $a>a_c$ and $1/D >$ 0 would be much more efficient. We refere to this method as "right side scoring", or right side search. An identical analysis can be made for a "left side scoring", in the case where the right side gets dispersed. Finally, from figure \ref{fig2} right panel, one can get the feeling of how difficult it is to recognize and efficiently detect a very old ancient family. Even with no objects other than the family members shown in this plot, it is hard to say for sure which is the best slope that fits the family. Therefore, even for a combined left and right search we can anticipate low detection efficiency.

\subsection{Need for Calibration}\label{calibration}

The difficulty in getting a good fit to a family V-shape as described in the previous section gets worse when the family is embedded in some background population. Recall that, the detection of a family by the V-shape method relies on the number of objects found above and below a nominal V, but within a predetermined sliver $dC~(dK)$ or $a_{\rm w}$. The combination $a_c$ and $K$ that returns the highest value of $N_{in}^2/N_{out}$ is identified as the center ($a_c$) and slope ($K$) of the family found. However, embedding the family in a background of unrelated objects can make even a well defined and visible V shaped family like the one presented in figures \ref{fig1} and \ref{fig2} left panel to become fuzzy (see \citealp{bolin2017} and Fig. \ref{fig7} in this work for examples). Therefore the maximum score can mislead the real center ($a_c$) and slope ($K$) of the family, leading to an incorrect or false detection.

Another complicating factor that could mislead the detection of the real center and slope of the family is the so-called {\it edge effect}. As anticipated in the previous section, {\it edge effect} is mostly a result of having a small number of objects outside the V related to those inside the V (where $N_{out} \rightarrow$ 0 would lead to $N_{in}^2/N_{out} \rightarrow \infty$). Therefore, edge effect could result as a strong signal in a K vs $a_c$ diagram  \citep[commonly used in many previous works as a tool to find family signals;][]{bolin2017,bolin2018b,bolin2018a,delbo2017,delbo2019}, even if no family was present. However, we have to point out that: 1) The {\it edge effect} is highly dependent on the searching method applied, i.e., whether both-sides or left/right-side are used. 2) The {\it edge effect} is also highly dependent on the choices of $a_{\rm w}$ and $dC$ (sliver width) made. 3) The {\it edge effect} is dependent regarding the range of slopes that we perform our search (i.e., steeper slopes tend to have the signal of the edge closer to the sample's vertical edge, because they are more vertical, than a shallower slope). Furthermore, it is plausible that a maximum in the $N_{in}^2/N_{out}$ due to an {\it edge efect} might not pass the so-called statistical test (i.e., the measured probability level at which we can reject (or not) the null hypothesis that the detected V-shape is created by random drawing from the observed size independent distribution of semimajor axes \citep{delbo2017,delbo2019}; not applied in the present work). We will come back to the issue of possible {\it edge effects} in section \ref{criteria}, when describing our criteria to identify a family signal, as well as in section \ref{application} when discussing practical application. 

Summarizing, in the real world the family search is done in parts of the asteroid belt that may contain both family members and background population. The asteroid belt itself could be entirely composed of members of unidentified families \citep{dermott2018}. These aspects of the real asteroid belt, along with the possibility of misleading signals due to edge effects, precisely reflect why we need to understand and characterize the searching tools. Therefore, we need to calibrate these methods in order to understand how efficient they are in finding families of different ages embedded in different levels of background (light, with small $N$; heavy, with large $N$).

In the following sections, we will proceed by creating and evolving a synthetic background and a series of synthetic families. Then, by knowing exactly what the center and slope for the synthetic families for different ages, we will embed them within different levels of random background objects and blindly search for these families.  Because we know a priori both center ($a_c$) and slope ($K$) of our synthetic families at all ages, we can directly compare the detection with those values. Therefore, by doing this several times we can determine how efficient the methods are as a function of the size of the family over the size of the background, age (or slope), and $dC~(dK)$, and $a_{\rm w}$.

\subsection{Synthetic Background}\label{sbkg}

We have to use a synthetic background because anywhere in the actual main belt there is a risk that any population of asteroids may also include an existing undiscovered family. We need to integrate a larger population of asteroids under the influence of the planets in order to capture the complex dynamics of the main belt to be sure not to overestimate the behavior of the algorithms by under-estimating the orbital complexity of the asteroid belt. Note that one possible scenario for the actual main belt is that the background is composed entirely of old families \citep{dermott2018}; precisely why we are building the synthetic background via direct integration.

To create such a synthetic background we start with the finding by \citet{tsirvoulis2018}. 
The aforementioned work, by removing all known asteroid families within the so-called pristine zone; 2.82 au $< a <$ 2.96 au and bracketed by the 5:2 and 7:3 MMRs with Jupiter; determined 
that a good representation of the SFD of the remaining background asteroids within that region follows a cumulative function $N(>D) \propto D^{-q}$ with $q =$ 1.43. The work by \citet{tsirvoulis2018} also concludes that a slope of $q=$ 1.43 is likely primordial. We thus considered such SFD and generated $N_{bkg}=$ 10,000 objects to constitute our synthetic background. Still, we anticipate that the choice of background SFD has only minor impacts in our methodology, as discussed in section \ref{conclusion}. We cut off the tail end of our SFD in objects of 2 km in diameter. These objects were randomly distributed within $a =$ [2.1-2.5] au, $\sin(i) =$ [0:0.35], and had their eccentricity such as $q>Q_{Mars}$, where $q=a(1.-e)$ is the perihelion of the asteroids and $Q_{Mars}=a_{Mars}(1.+e_{Mars})$ is the aphelion of Mars.

These objects were numerically integrated for 100 Myr under the influence of all solar system planets from Venus to Neptune, as well as under the influence of the Yarkovsky effect. We used the symplectic integrator known as swift rmvs3 \citep{levison1994} with a time step of 0.03 yr. To account for the Yarkovsky effect we modified the integrator by adding acceleration terms. We considered a simple diurnal $da/dt$ term in the equations of motion \citep{walsh2013,delbo2017}. We assumed the present day orbits of the planets as their initial conditions. Therefore, we expect to have been able to capture all the complex dynamics in the inner main belt region without losing a huge number of asteroids and decreasing $N$. Finally, after 100 Myr we averaged the last 10 Myr of evolution of the $a,e,i$ in order to estimate what we call a quasi-proper $a,e,i$ for all synthetic background objects.

\subsection{Synthetic Families}\label{sfam}

We need to generate and evolve synthetic families because it is of major importance for us to know exactly the age (slope) and center of the family over time, as well as the number of family members and their distribution in $1/D$ and quasi-proper $a,e,i$. We need to know precisely these quantities so that we can characterize the efficiency of the searching method.

Similar to what was done for the synthetic background, in order to create a synthetic family we chose a large number of objects, $N_{fam} =$ 3000, and a family SFD.  We opted to use an SFD similar to that of the Erigone family ($q \sim$ 3.5); because Erigone is a well studied family and it is also relatively young (age $\sim$200 Myr; \citealp{spoto2015} using a V-shape fit, $\sim$300 Myr \citealp{vokrouhlicky2006} using HCM and YORP cycles). The initial dispersion of the synthetic family in $a~vs~1/D$ space was determined by a simple relation where the maximum distance in au from the center of the family that a object could be initially placed was ($- 2v_{init}/\eta$), and the actual distance was randomly selected between $a_c$ and ($- 2v_{init}/\eta$), where $\eta$ is the mean motion of an asteroid at the family's center location. This initial velocity dispersion has the same size dependence as Yarkovsky drift, which is a simple 1/$D$ dependence, so a family is initially a V-shape in $a~vs~1/D,H$ and its width is simply related to $v_{init}$. We adopted $v_{init} =$ 20 m/s. 

We should, however, point out that our choice of $v_{init}$ does not represent the correct initial dispersion of very large families, created by the disruption of large targets. Very large families could have $v_{init} \approx$ 100 m/s, or generally it is thought that initial velocity dispersion is similar to target body escape speed. This means that the initial dispersion in semimajor axis of the family members could span over the entire inner main belt. 

Such a large initial dispersion in semimajor axis could be misinterpreted as family drift over time and be interpreted as an age of Gyr according to the slope that would be detected by the V-shape method. Thus, it is typical to factor in estimates for a families parent body size to separate out the initial dispersion and that caused by drift. However, here, to make the slope detection consistent with the age of the family, we chose $v_{init} =$ 20 m/s. Thus, our synthetic family would have an initial age smaller than 10 Myr, and we can then consistently relate our detection (slope, K) with the age of our synthetic family, as it evolves. Still, we should ratify that the goal of the present work is to characterize the efficiency of the detection method. A full characterization of the family would demand additional steps \citep[e.g., confirmation with clustering methods (HCM), statistical analyses, comparison with spectroscopic, SFD, etc...;][]{bendjoya2002,nesvorny2015,delbo2019}.

The obliquity of each object is selected randomly between 0-180 degrees, and does not change throughout the simulation (although we know from \citet{vokrouhlicky2006} that YORP cycles can change asteroid's obliquities, we want to keep the evolution as simple as possible). This scales the drift of an object, where it is also scaled by the diameter, density and distance relative to the values established for B-type asteroid Bennu \citep{chesley2014}.  

\subsection{Family evolution -- dynamics}\label{famevodin}

\begin{figure*}
 	\centering
 		\includegraphics[scale=0.29]{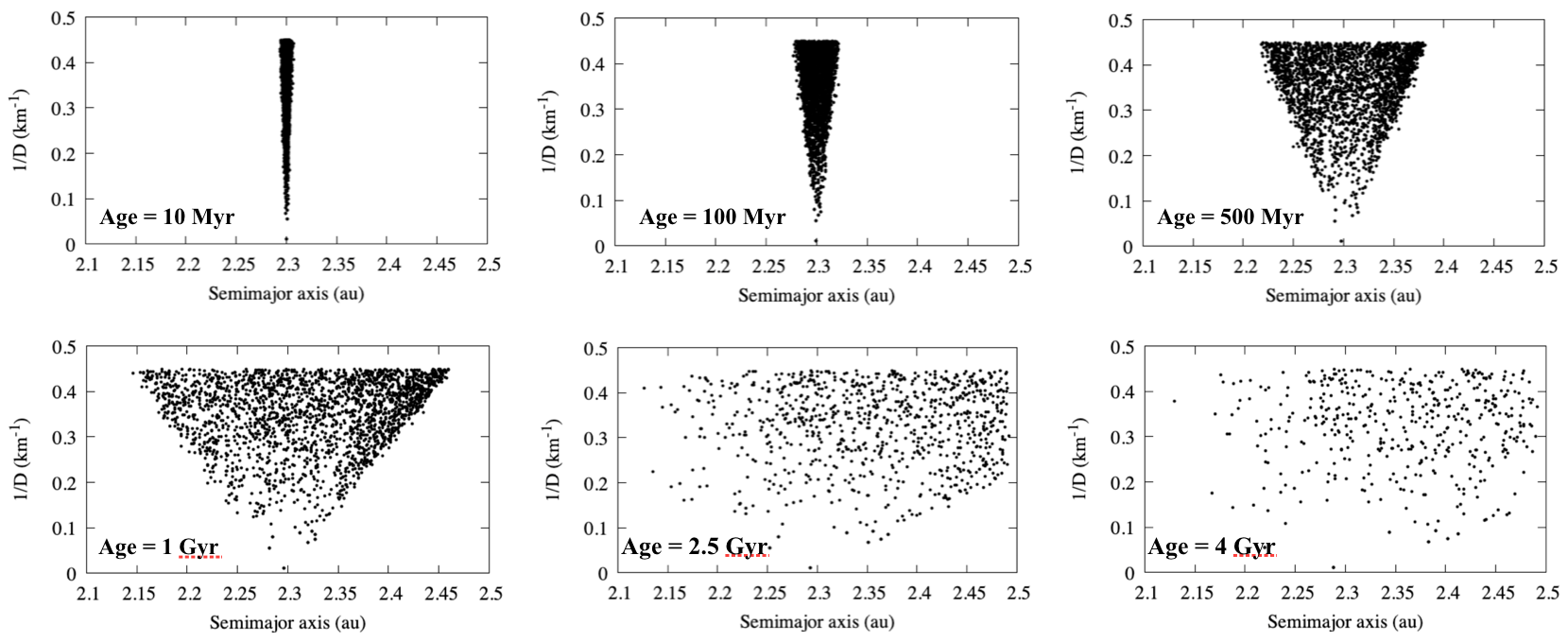}
 	\caption{Temporal evolution of a synthetic family perturbed by all solar system planets from Venus to Neptune and Yarkovsky drift over 4.5 Gyr time span.}
 	\label{fig3}
 \end{figure*}

Once created, we evolve the synthetic family for 4.5 Gyr in the same way we did for the synthetic background in section \ref{sbkg}, i.e., under the influence of the gravitational perturbation of Venus, Earth, Mars, ~Jupiter, ~Saturn, ~Uranus, and Neptune, as well as the non gravitational Yarkovsky drift, while also using the same integration method. Figure \ref{fig3} shows snapshots of the evolution of our synthetic family in the proper $a$ (au) $vs$ $1/D~(\rm km^{-1})$ space. In each panel of figure \ref{fig3}, we have averaged over the last 10 Myr of evolution of the family so that we could get the quasi-proper elements for the family members.

As shown in figure \ref{fig3}, the opening of the V shape of the family enlarges due to the Yarkovsky drift imposed on family members' semimajor axes. As well, when the family members evolve through MMRs they either get lost or dispersed, usually changing the V shape form of the whole family. Up to about 1 Gyr the family still present an almost perfect V-shape. On the other hand, in the case of a family of 4 Gyr of perturbed evolution, the family already looks almost unrecognizable and more similar to what we would expect for a random background population.

\begin{figure*}
 	\centering
 		\includegraphics[scale=0.29]{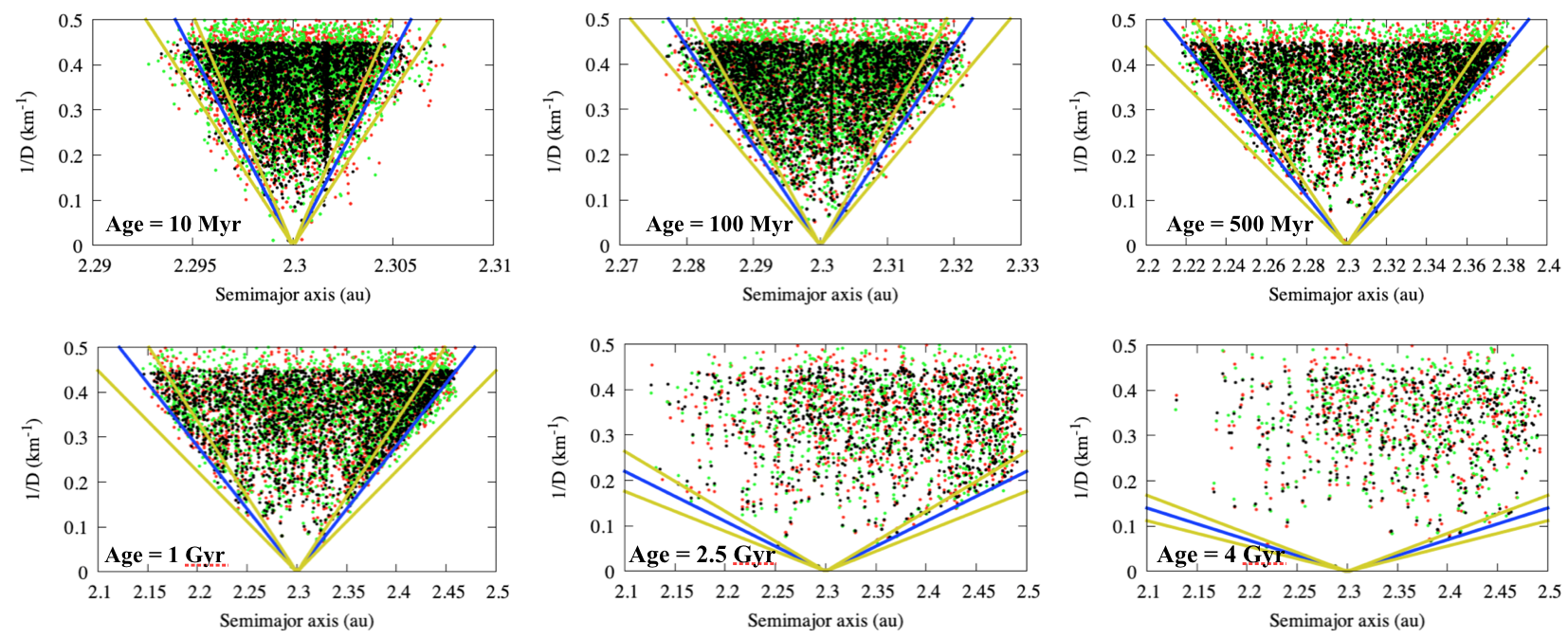}
 	\caption{Same as figure \ref{fig3} but showing in red and green two new synthetic families after applying uncertainties of $\pm$10-15 \% in the nominal (black) diameter and $\pm \sigma_a$ (see main text) in the nominal quasi-proper $a$. The blue line represents the best V fit for each different age within an uncertainty of $\pm$20\% (yellow lines).}
 	\label{fig4}
 \end{figure*}

In the evolution shown in figure \ref{fig3} all objects have the SFD described in section \ref{sfam} and create a real V-shape distribution. However, in reality, astronomical observations of each asteroids brightness, converted to a size, will not be perfect and even very young and abundant V-shapes may not be as sharp. To account for this we included in our data some uncertainty in both diameter and quasi-proper $a$. The uncertainty in the diameter was assumed to $\pm$10-15\% of the original diameter \citep{harris2006,masiero2018}. For the quasi-proper $a$ we added an uncertainty factor of $\pm \sigma_a$ to the averaged value of quasi-proper $a$, where $\sigma_a$ is the standard deviation from the averaged quasi-proper $a$ that comes from the simulation. We did this 99 times so to generate 100 different, but similar synthetic distributions of our synthetic families. Our new 99 synthetic families created are no longer perfect V-shapes (see figure \ref{fig4}). Once embedded in the background population (section \ref{fambkg}), the transition of these families' V edge to the background population will be smoother than that presented in figure \ref{fig7}, where we plotted our nominal and perfectly V-shaped synthetic family as reference. Therefore, although not necessarily ideal, these additional families with smoother transition to the background population should roughly represent the case where the core of the V-shape distribution would represent the family and the blurry edge of the V would potentially represent some or all of the family's halo \citep{parker2008,nesvorny2015,broz2019}. Still, we call attention to the fact that halos are mostly visible in the ($e,i$) plane.

Figure \ref{fig4} shows how the families from figure \ref{fig3} will be represented after the uncertainties are applied. Also in this figure we show which is the best slope (blue) for a V fit for each age within an uncertainty of $\pm$20\% (yellow). Although 99 additional families were created, we only plotted two of them along with the nominal synthetic family from figure \ref{fig3} to improve visualization. All these 100 families per age will be used as templates for characterizing our V-shape search.

The slopes shown in figure \ref{fig4} will be the target for our searching method. Any detection returning a slope within the yellow limits will be considered a good detection (see section \ref{criteria} for more details in our detection criteria). Although not entirely shown, our family age (slope) sample is composed by 10, 30, 50, 100, 300, 500 Myr, and 1.0, 1.5, 2.0, 2.5, 3.0, 3.5, 4.0, and 4.5 Gyr. For all these ages we fitted an optimal slope within $\pm$20\% uncertainty to be used as reference values, as we did in figure \ref{fig4}.

Especially for the case of old families, where the V-shape due to Yarkovsky effect is significantly wider than the V-shape due to the initial velocity field, the age of a family becomes highly correlated with the Yarkovsky drift rate $da/dt$ [age = $f(da/dt)$], which is a function of size, density, thermal inertia, obliquity, period of rotation, albedo and distance from the Sun. Therefore, from now on, when discussing detection, we will no longer refer to the age of the family, but rather to the slope that the family has. This is an important distinction because although we want to characterize how efficient a family of a given age could be detected as a function of the ratio between the density of family members with respect to that of the local background, $dK$ ($dC$) and $a_{\rm w}$, the technique searches primarily for slopes. With that said, characterizing slope detection is more consistent and coherent. Besides, especially because the method search for slopes, the characterization is straight forward. Once the slope is found, by estimating the Yarkovsky drift rate $da/dt$ one can infer the age for the family.

\subsection{Family evolution -- collision/depletion}\label{famevocoll}

It is well known that families collisionally evolve over time \citep{bottke2005a,bottke2005b}. Therefore, the number of objects within a family (especially for small diameter members) can be decreased due to collisional evolution and also due to dynamical depletion. It is essential to also understand how collisional effects should affect the size (number) and density (number/area) of a family over time.

\begin{figure*}
 	\centering
 		\includegraphics[scale=0.29]{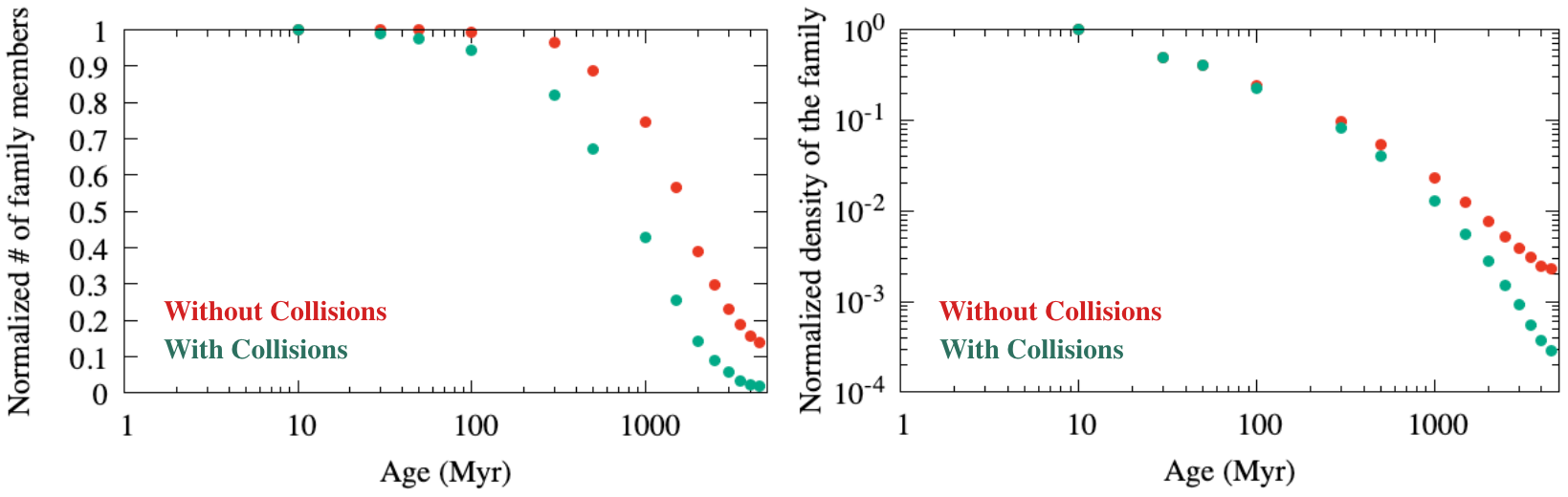}
 	\caption{Normalized number of family members (left) and density of the family (number of members within the area enclosed by the V-shape; right), as a function of age for the cases when collisional evolution is considered (green) and when it is not (red). The area enclosed by the V is defined as: $V_{area}={\rm max}(1/D)(a_{max}-a_{min})-\int_{a_{min}}^{a_{max}}K|a-a_c|da$, where $a_{min}$ and $a_{max}$ are the semimajor axes of the leftmost and rightmost family member, with $a_c$ the center of the family and $K$ and $D$ the slope and diameter.}
 	\label{fig5}
 \end{figure*}

Figure \ref{fig5} shows results from the simulation of a synthetic family over time with its total number of bodies output at the sample of ages (10, 30, 50, 100, 300, 500 Myr, and 1.0, 1.5, 2.0, 2.5, 3.0, 3.5, 4.0, and 4.5 Gyr) when considering and not considering the collisional algorithm from \citet{bottke2005a,bottke2005b}. The \citet{bottke2005a,bottke2005b} results are used in a probabilistic manner, where each asteroid was removed as a function of its current lifespan compared to its expected collisional lifetime. Simply, at each time step of the simulation a random number is selected and used to determine whether each asteroid of given collisional lifetime would be destroyed. As the life expectancy of the asteroid is a function of the diameter only, this causes the smaller asteroid to be eliminated faster than the larger ones.

As expected and already observed in many of the previous figures, the number of objects decrease over time simply from the dynamics of loss via resonances. In addition, collisional evolution decreases the population further. Figure \ref{fig5} shows an increase of loss within the families when considering collisions that is primarily due to the depletion of the smallest members of the family. This is intuitive and observed in very old families that are mainly composed by larger objects \citep{delbo2017,delbo2019}. 

Regardless of considering or not collisional evolution, the density of a family in the quasi-proper $a~vs~1/D$ space decreases to $\sim$90\% within the first 100-300 Myr (figure \ref{fig5} right panel). This is mostly related to spreading in semimajor axis, which increases the area inside the V very fast. This spreading of the area the family covers at first overwhelms any losses due to collisions, as the number of objects lost within the first 100-300 Myr is only 10-20\% (figure \ref{fig5} left panel).

\begin{figure}
 	\centering
 		\includegraphics[width=\columnwidth]{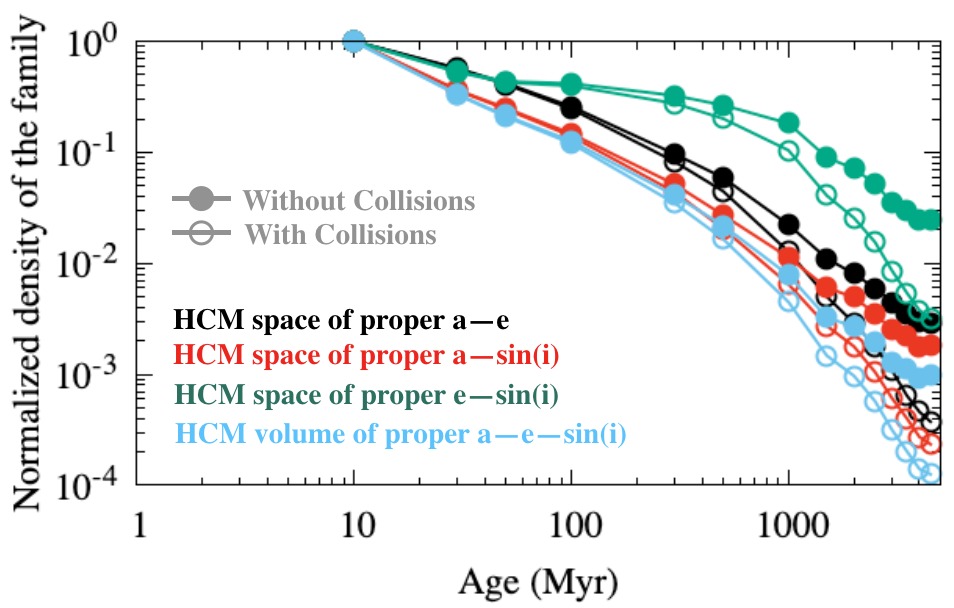}
 	\caption{Normalized density of the family (number of members within the rectangular area enclosed by the minimum and maximum value of each quantity distribution) in the HCM proper elements space, considering collisional evolution (open circles) or not (filled circles). Black: HCM space of proper $a$ $vs$ proper $e$. Red: HCM space of proper $a$ $vs$ proper $\sin(i)$. Green: HCM space of proper $e$ $vs$ proper $\sin(i)$. Blue: HCM volume of proper $a$ $vs$ proper $e$ $vs$ proper $\sin(i)$.}
 	\label{fig6}
 \end{figure}

Similar to figure \ref{fig5}, in figure \ref{fig6} we show how the density of the family as a function of age changes in the proper elements of the HCM.

Figure \ref{fig6} reinforces the observation that the spreading in proper $a$ is the biggest factor for the decrease in family's density. Also from figure \ref{fig6} we see that spreading in proper $e$ and proper $\sin(i)$ are not so large for the synthetic family. These conclusions come from the fact that the blue curve is very similar to that shown in figure \ref{fig5} right, so as the red and black. Also this explain why the green curve falls off the other curves. It implies much smaller dispersion within proper $e$ and proper $\sin(i)$.

As a final note, it is expected from the results presented in figures \ref{fig5} and \ref{fig6} that our studied synthetic families should in principle be considered more depleted in small objects than we in fact consider. However, this should not be seen as a negative point because, 1) the V-shape method search for the edges of the V, which does not necessarily need to account for the small objects that would be supposedly lost (see figure \ref{fig1}) and 2) it is not clear whether or not the small population would be refilled over time by collisional evolution of the larger members, and if they were they would stay within the boundaries of the initial expanding V-shape. Keeping these two points in mind we then decided to continue considering all the members of our synthetic families in the following analyses and not artificially removing some of them.

\subsection{Combining the Families and Background for V-shape identification testing}\label{fambkg}

Before we describe in detail how we defined our detection criteria and start the characterization tests,  we need to describe how we combined synthetic family and synthetic background. This is important because this will be the test bed for all of our conclusions.

\begin{figure*}
 	\centering
 		\includegraphics[scale=0.35]{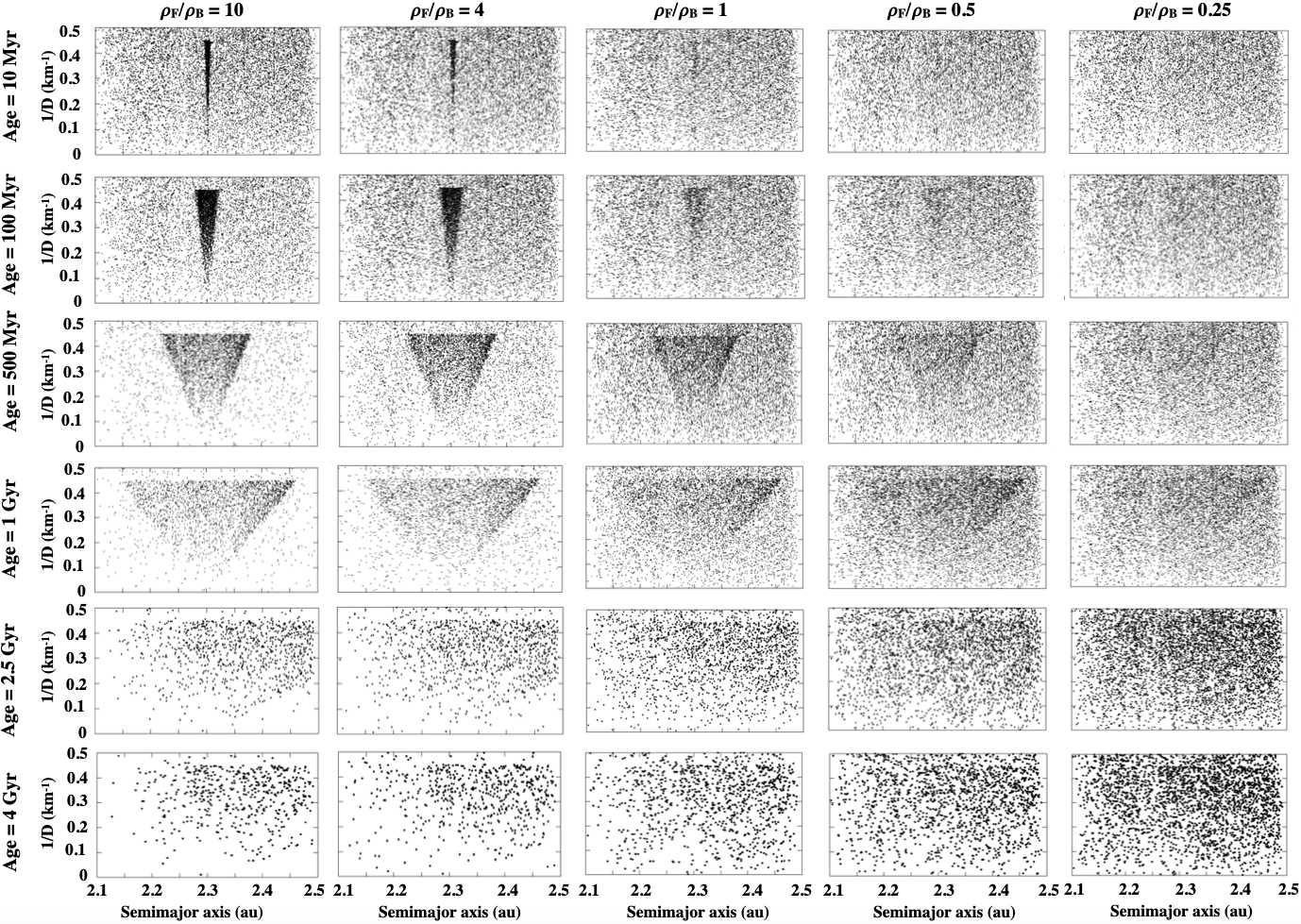}
 	\caption{A representation of synthetic families embedded in synthetic backgrounds for different signal-to-noise levels ($\rho_F/\rho_B =$ 10, 4, 1, 0.5, 0.25) and for different family ages (10, 100, 500 Myr and 1, 2.5, 4 Gyrs). The y-axes show $1/D$ up to 0.5 $\rm km^{-1}$ while the x-axes show semimajor axis from 2.1 au to 2.5 au. Only our nominal, perfectly V-shaped, family is shown to improve visualization. All other 99 additional, non-perfectly V-shaped, families generated as described in section \ref{famevodin}, with blurrier edges, would present a smoother transition to the background population.}
 	\label{fig7}
 \end{figure*}

As discussed in section \ref{calibration}, even a very well defined V shaped family can become fuzzy when embedded in a background population. The level of fuzzyness, however, depends on the ratio of the density of the family over the density of the background. We define the density of the family $\rho_F$ as in the previous section ($\rho_F = N_{fam}/V_{area}$, see figure \ref{fig5}). The density of the background is defined as $\rho_B = N_{bkg}/A_{sqr}$, where $N_{bkg}$ is the number of background objects and $A_{sqr}$ the constant square area $A_{sqr} = (a_{bkgmax} - a_{bkgmin})(1/D_{bkgmax}) = (2.5-2.1)0.5 = 0.2$ au ${\rm km^{-1}}$ of our background space. We then define different levels of family and background such as the ratio $\rho_F/\rho_B$ will determine our signal-to-noise level. For that, we randomly select a sample of objects within $N_{fam}$ and within $N_{bkg}$ in order to have a desired signal-to-noise. In other words, from our available sample ($N_{fam},N_{bkg}$), we either decrease one or the other so we can change $\rho_F/\rho_B$ accordingly. A visualization of the above description can be found in figure \ref{fig7}, for our nominal case.

We have to point out a caveat for our signal-to-noise definition. This caveat is: because in the real world we do not know a priori what is family and what is not within our sample, our density calculation might not be directly applicable. However, we also have to remark that, although not directly applicable to the real situation, as the goal of this work is to well characterize the V-Shape searching method under general conditions, the methodology is well suited for the test bed cases we need.

The scenario presented in figure \ref{fig7} exemplifies the sample where we will apply our searching methods. Although we did not show all cases in figure \ref{fig7} due to visualization constraints, we embedded every one of the 100 synthetic families for each age in a random selection of background objects as shown in figure \ref{fig7}. Figure \ref{fig7} shows examples of our nominal and perfectly V-shaped family at different ages fading out when the signal-to-noise level decreases from 10 to 0.25 (in the practical case we considered 8 different $\rho_F/\rho_B$ levels, say 0.25, 0.5, 1, 2, 4, 6, 8, and 10). The fading effect is much stronger for the non-perfectly V-shaped families, with blurry edges, created as described in section \ref{famevodin} (thus not shown to improve visualization). In total we have created 100 background plus family samples for each of the 14 different ages within 8 $\rho_F/\rho_B$ levels. Therefore, in the end we have a total of 11,200 family plus background samples. Also, as one can see, our family plus background sample covers all possible situations where families are very likely to be found (leftmost panels, especially those on the top) as well as those where we think to be closer to the real case within the main asteroid belt (rightmost panel), where we expect the method to struggle a bit.

\subsection{Family detection criteria}\label{criteria}

 \begin{figure*}
 	\centering
 		\includegraphics[scale=0.35]{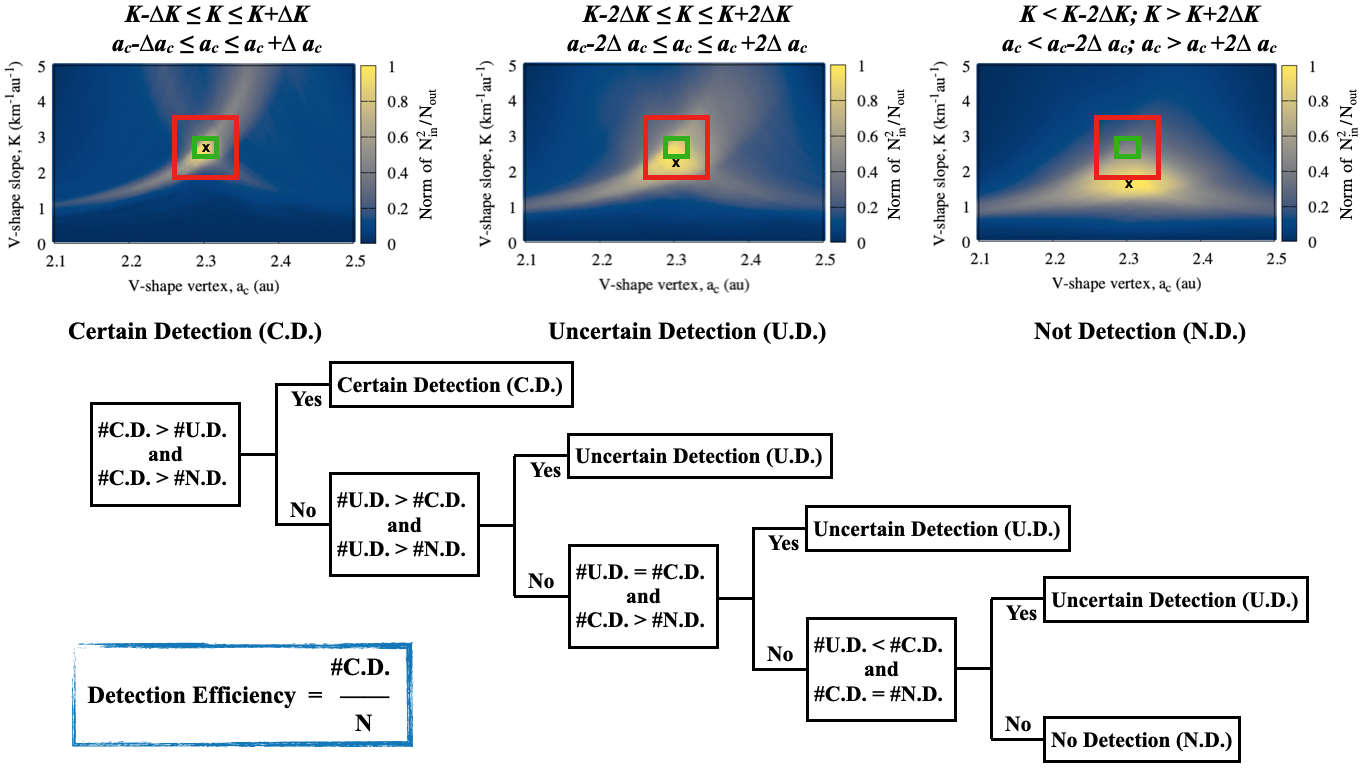}
 	\caption{Schematic diagram of our family detection criteria. Top: Three different colored score maps showing examples of possible three different signals that can be found when a search in the ($a_c,K$) space is performed. Top left represents a case where $K-\Delta K \leq K \leq K+\Delta K$ and $a_c-\Delta a_c \leq a_c \leq a_c+\Delta a_c$ (green box). We designate this case as a Certain Detection (C.D.). Similarly, top center panel represent what we designate Uncertain Detection (U.D.), with the highest score (black cross) falling between the green and red boxes (the limits of the red box are defined on top of the panel). Top right: a case outside the red box, representing what we call Not Detection (N.D.). Below the scoring maps, a logical chain showing the rules applied to determine whether, after all 100 searches for signals of a given synthetic family are performed, will result in C.D., U.D., or N.D. (recall that each search will result in a C.D., U.D., or N.D. score, and therefore after 100 scores of a given family, for the total numbers of C.D, U.D., or N.D. -- \# C.D., \# U.D., \# N.D. -- we apply our logical chain; more detail in the main text). The blue box in the bottom left highlights how we computed the Detection Efficiency (D.E.).}
 	\label{fig8}
 \end{figure*}

An automated detection criteria is necessary to analyze all 11,200 cases studied here. Moreover, we need to automate both the search method and the way we check for detection. The requirement of automation becomes even more clear considering that as we also intend to precisely measure how the methods are sensitive to $dC$ ($dK$) and $a_{\rm w}$ \citep{bolin2017}, for each of the 11,200 synthetic sets we will run 250 different $dC$ slivers and 150 different $a_{\rm w}$ slivers (figure \ref{fig1}). For the $dC/dK$-$method$ we varied $dC$ from $2\times 10^{-7}$ au to $5\times 10^{-5}$ au with 250 increments of $2\times 10^{-7}$ au, where $K\pm \Delta K=\sqrt{pv}/[D0(C\pm dC)]$. For the $a_{\rm w}$-$method$ we varied $a_{\rm w}$ from 0.001 au to 0.1 au with 100 increments of 0.001 au, and after from 0.11 au to 0.6 au with 50 increments of 0.01 au. Therefore, in the end we will have performed 2,800,000 cases for the $dC/dK$-$method$ and 1,680,000 cases for the $a_{\rm w}$-$method$, which will produce enough statistics to characterize the methods.

We automate the search method by the following prescription. We run a loop over all different 14 ages in our set (10, 30, 50, 100, 300, 500 Myr, and 1.0, 1.5, 2.0, 2.5, 3.0, 3.5, 4.0, and 4.5 Gyr). Inside this loop we have a new loop over all 8 different signals to noise levels ($\rho_F/\rho_B$ = 10, 8, 6, 4, 2, 1, 0.5, 0.25). Then for each pair age and $\rho_F/\rho_B$, assuming a fixed values of $dC$ or $a_{\rm w}$ slivers we run another loop over all 100 synthetic family plus background created in the previous section. In other words, consider a family of age X from our sample. We then embed this family in a $\rho_F/\rho_B$ = Y. Once the age and $\rho_F/\rho_B$ was defined, we fix a value of $dC=dC_1$ or $a_{\rm w}=a_{\rm w1}$ sliver and apply our detection methods for each of the 100 synthetic families of that age (those created in section \ref{famevodin}, e.g. figure \ref{fig4}) embedded in the corresponding synthetic background level as shown in figure \ref{fig7}. Each of the 100 searches are performed within the interval of slope $K=[K_{min}=0.1; K_{max}=100]~\rm au^{-1}km^{-1}$ and $a_c=[a_{cmin}=2.1; a_{cmax}=2.5]~au$. Also, each of the 100 searches will result in a score map as shown by the color maps in figure \ref{fig8} top panels (truncated at $K=$ 5 $\rm au ^{-1}km^{-1}$ for better visualization). This procedure is repeated for all ages and $\rho_F/\rho_B$ combinations, as well as for all different values of $dC$ and $a_{\rm w}$ slivers for a given (age,$\rho_F/\rho_B$) combination.

We automate the way we check for detection assuming pre-defined intervals of slope ($K$) and center ($a_c$) and comparing with the results from our score maps (figure \ref{fig8} top panels). 
The nominal values for $K$ and $a_c$ are known and were obtained from our nominal simulations (section \ref{famevodin}, figures \ref{fig3} and \ref{fig4}). The nominal values to compare our results with are, $a_c$ = 2.3 au in all ages, and $K (\rm au^{-1}km^{-1})$ = 85, 45, 37, 22, 9, 5.5, 2.8, 1.9, 1.5, 1.1, 0.95, 0.82, 0.7, 0.65 for ages 10, 30, 50, 100, 300, 500 Myr, and 1.0, 1.5, 2.0, 2.5, 3.0, 3.5, 4.0, and 4.5 Gyr respectively. Therefore, for each of the 100 score maps obtained for a certain pair age and $\rho_F/\rho_B$ with fixed values for $dC=dC_1$ or $a_{\rm w} = a_{\rm w1}$ slivers from the automated search described in the previous paragraph we record the maximum normalized values for $N_{in}^2/N_{out}$ and the associated pair $(K,a_c)_{best}$. We then compare these $(K,a_c)_{best}$ pairs with our nominal $(K,a_c)$ pairs within a tolerance interval of $\pm x\Delta K$ and $\pm x\Delta a_c$ (where $\Delta K=\pm 20\%$ of the nominal $K$ and $\Delta a_c = \pm 0.01$ au).
If according to our tolerance definition $(K,a_c)_{best}$ is within $x\leq 1$ we flag this detection as Certain Detection (C.D.), figure \ref{fig8} top left panel. If $(K,a_c)_{best}$ lies within $1 < x \leq 2$ we flag this detection as Uncertain Detection (U.D.), figure \ref{fig8} top center panel. Lastly, if $x>2$ we flag this detection as Not Detection (N.D.), figure \ref{fig8} top right panel. Thus, for the total number of C.D., U.D., and N.D., we run a logical chain as shown in figure \ref{fig8} right below the colored score maps. This way we determine whether the final result will be C.D., U.D., or N.D.. 

In order to clarify the method, let us assume the same example from the previous paragraph, say, a family of age X and a $\rho_F/\rho_B$ = Y assuming a fixed value for $dC = dC_1$ or $a_{\rm w} = a_{\rm w1}$ slivers. Each of the 100 searches performed for the synthetic families in this situation by the automated search will result in a pair $(K,a_c)_{best}$. Therefore, there will be 100 $(K,a_c)_{best}$ pairs and each of these pairs will have a C.D., U.D., or N.D. flag associated. If, for example, in the end we have \# C.D. = 40, \# U.D. = 50, and \# N.D. = 10, by applying the logical chain from figure \ref{fig8} the ultimate result will be Uncertain Detection (U.D.). Once again, this procedure is repeated for all ages and $\rho_F/\rho_B$ combinations, as well as for all different values of $dC$ and $a_{\rm w}$ slivers for a given (age,$\rho_F/\rho_B$) combination.

It is now important to discuss, once again, the context of {\it edge effects}. The three points discussed in section \ref{calibration} regarding {\it edge effects} lead to the following common results according to our methodology: a distorted map (with multiple centers), a null map (with basically no signal at all), or a very fuzzy signal that expands over a large range of semimajor axis. Therefore, we conclude that {\it edge effects} are not a problem for our characterization methodology (as explained in this section), which compares the strongest signal found with an a priori known center and slope for a given family age (recall that, in our experiments, we know where our synthetic family is at all times). This means that, if a strong signal is found as a result of an {\it edge effect}, our methodology would return 
Not Detection (N.D.).

Finally, we define the quantity Detection Efficiency (D.E.) as \# C.D./$N_{cases}$ = \# C.D./100, with $N_{cases}$ the number of cases per pair (age,$\rho_F/\rho_B$) analysed (100 in this work). It is worth noticing that even a family flagged as U.D. or N.D. may have an associated quantity D.E. different than zero. The reason behind is, let us suppose within our 100 families, for a value of $dC = dC_1$ or $a_{\rm w} = a_{\rm w1}$ we have \# C.D. = 10, \# U.D. = 30., and \# N.D. = 60. In this case, this family will be flagged as N.D. according to our logical chain in figure \ref{fig8}, but it will still have a non zero D.E. = 0.1 or 10\% efficiency. Therefore, for the sake of clarity, in the following we will only discuss cases with D.E. $\geq$ 0.5 or 50\% efficiency.

\section{Results}\label{results}

In this section we discuss the main findings from the described simulations. First, we show how the detection criteria depends on the choice of $a_{\rm w}$ ($dC$ or $dK$) for different signal-to-noise levels and slope. Later we focus the discussion only on cases that were flagged as Certain Detection (C.D.). This is done to provide a general view of the relationship that $a_{\rm w}$ ($dC$ or $dK$) has with slopes. In all cases we also present Detection Efficiency (or likelihood for detection).

\subsection{Detection criteria as a function of width}\label{detcriteria}

 \begin{figure*}
 	\centering
 		\includegraphics[scale=0.35]{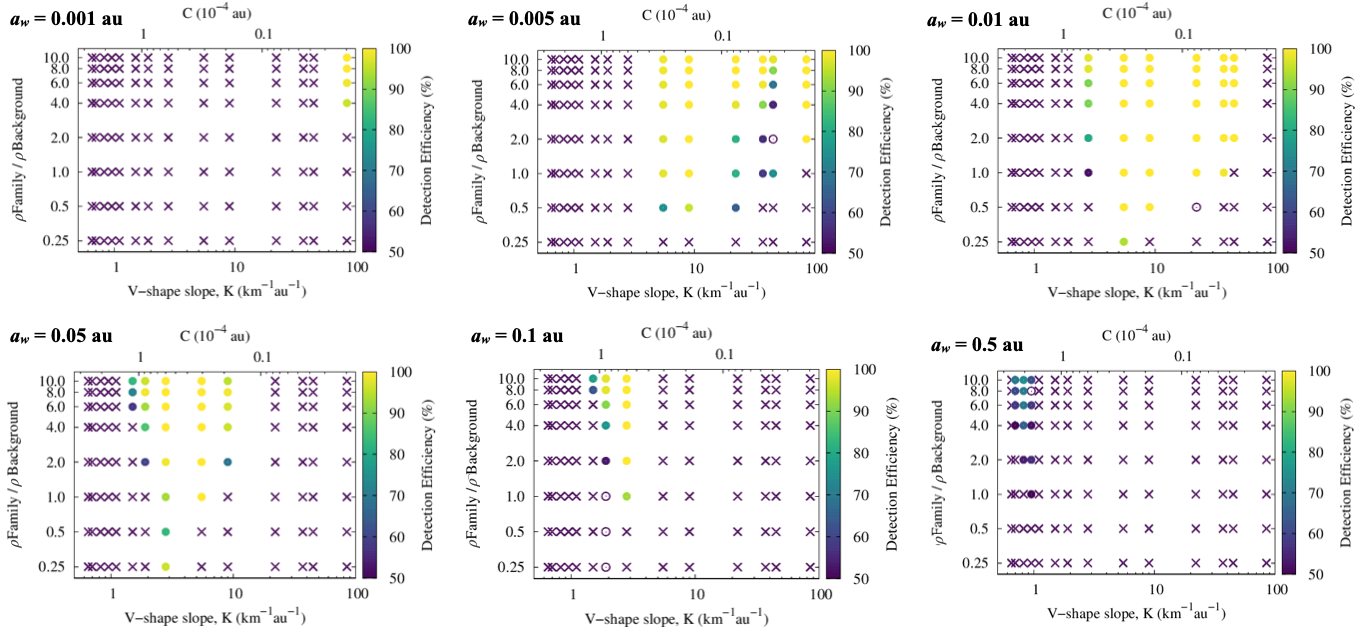}
 	\caption{Detection efficiency as a function of slope (age) and density of family to background for different values of $a_{\rm w}$ width considered (labeled as $a_{\rm w}$ = 0.001 au, 0.005 au, 0.01 au, top panels, and $a_{\rm w}$ = 0.05 au, 0.1 au, 0.5 au, bottom panels, all from left to right). Filled circles represent the cases flagged as Certain Detection. Open circles represent Uncertain Detection. Crosses represent Not Detection. The top of each panel also brings the corresponding values of the characteristic value $C$ \citep{vokrouhlicky2006}.}
 	\label{fig9}
 \end{figure*}

After performing all 2,800,000 cases for the $dC/dK$-$method$ and 1,680,000 cases for the $a_{\rm w}$-$method$ as described in the previous section, we show in figure \ref{fig9} when a family of a given slope (age) can be found (C.D.), cannot be found (N.D.) or is possible to be found (U.D.) as a function of $\rho_F/\rho_B$ (see section \ref{fambkg} for a detailed description of how we defined the different levels of $\rho_F/\rho_B$). For each flag C.D. and U.D. we also show (figure \ref{fig9}) the detection efficiency (D.E. = \# C.D./100), i.e. the likelihood for finding, of that given family. Note that the D.E. scale shows only values of D.E. $\geq$ 50\% once everything with D.E. $<$ 50\% is more likely to not be detected.

The results, compiled in figure \ref{fig9}, show some intuitive things, such as that both young and old families are more efficiently found against a lower density background (increasing efficiencies moving up each plot). It also shows that for each slope considered there is a clear optimal $a_{\rm w}$ width sliver where detection efficiency is maximized. 

Although only showed for $a_{\rm w}$ (when both side scoring was considered, see section \ref{awK} and \ref{awKlr}), similar results are observed in the $dC$ ($dK$) space and when a right side scoring is considered (recall that our synthetic family evolves to a right side family at older ages, section \ref{famevodin}). The results are very sensitive to the choice of the width of the sliver considered in the search (as previously noticed by \citealp{bolin2017}). In addition, the panels from figure \ref{fig9} give us some notion that the efficiency and likelihood for detecting the signal of a given slope vary in such a way that, the width of the sliver should increase as the slope decreases. Moreover, the results from figure \ref{fig9} also tell us that independently of searching for both or one side of the V, and even more surprisingly some times independently of the signal-to-noise level (case of moderate slopes, mid-age families), almost all family slopes have an associated detection efficiency larger than 50\% for an optimal value $a_{\rm w}$ or $dC$ of the sliver considered. This efficiency, however, is not equal for all scenarios. Therefore, to more accurately determine the optimal width of the sliver as a function of the slope to be found for each of the different searching methods, we dedicate the following sections.

\subsection{$a_{\rm w}$ as function of K; both side scoring in the $a_{\rm w}$-$method$}\label{awK}

 \begin{figure*}
 	\centering
 		\includegraphics[scale=0.35]{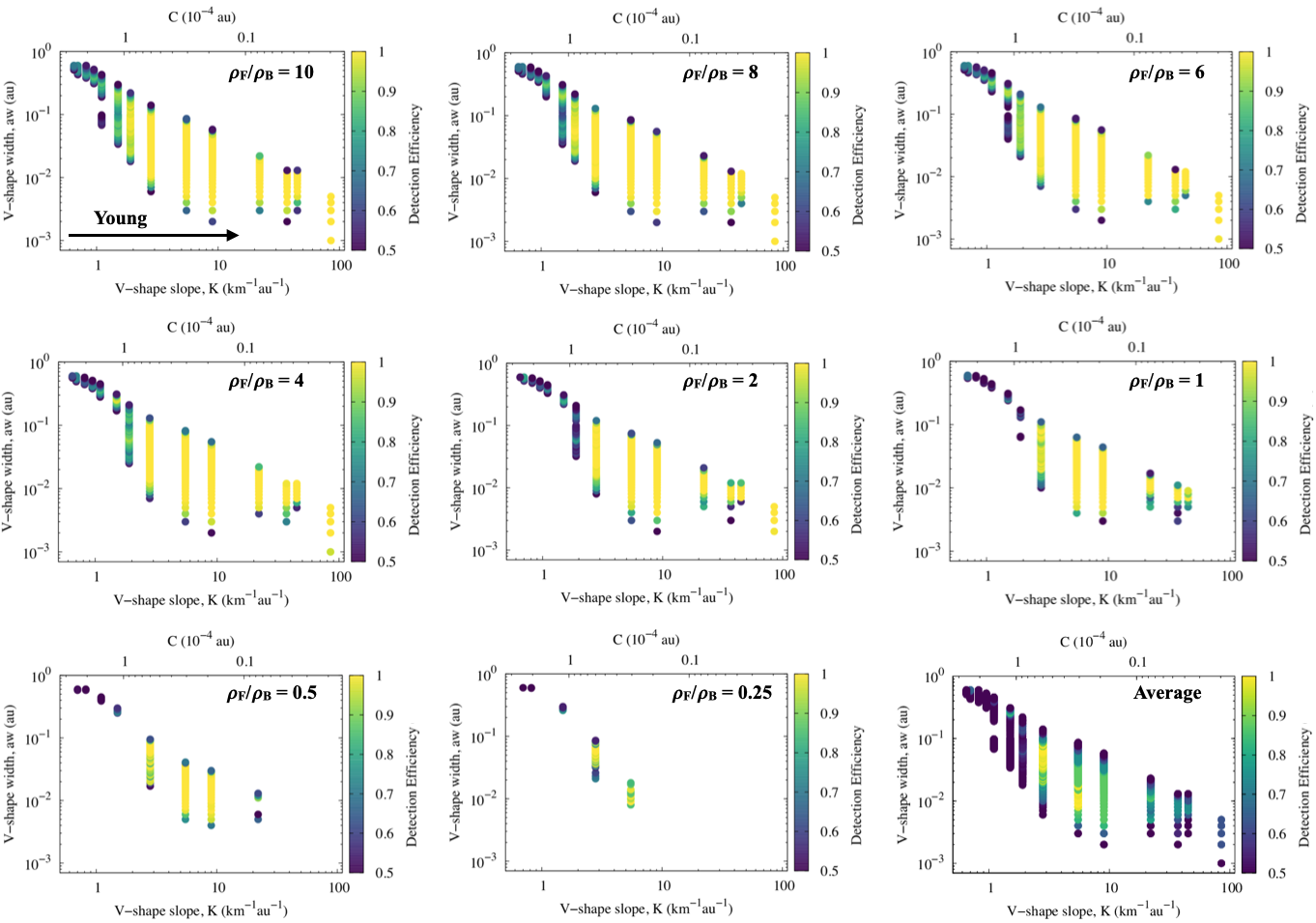}
 	\caption{Detection efficiency for $a_{\rm w}$ as a function of $K$ within $a_{\rm w}$-$method$ both side scoring.}
 	\label{fig10}
 \end{figure*}

There is a clear pattern of $a_{\rm w}$ that are optimal for detecting families of different ages. This pattern becomes evident when plotting as a function of the slope. Larger widths $a_{\rm w}$ are necessary for fitting older families and smaller widths for younger families (figure \ref{fig10} shows how the width $a_{\rm w}$ relates to the slope $K$ of a family within the $a_{\rm w}$-$method$). This clear pattern emerges for all values of signal-to-noise, or family density to background density, shown in figure \ref{fig10} as the detection efficiency for all pairs ($a_{\rm w},K$) that returned D.E. $\geq$ 0.5. Despite some small differences within the panels in figure \ref{fig10}, the shape of the curves that relate $a_{\rm w}$ with $K$ and the range of $a_{\rm w}$ widths that return D.E. $\geq$ 0.5 for a given slope $K$ are remarkably similar in all panels. Of course in very dense backgrounds, $\rho_F/\rho_B <$ 1, detection becomes less and less likely, as one can see in figure \ref{fig10} (see for example the case for $\rho_F/\rho_B =$ 0.25 which indicates that young families are likely not detectable by the $a_{\rm w}$-$method$ under this situation). However, within the few cases where detection was successful for such small signals to noise levels, the shape of the $a_{\rm w} = f(K)$ curve and the range of widths with D.E. $\geq$ 0.5 follows the same pattern as for the higher signal-to-noise level cases.

The absence of randomness in scattering of points in these panels lead us to the conclusion that regardless of size or density that a given family may have with respect to a background population, there is an optimal range of $a_{\rm w}$ widths that perform best with this method, and thus can more effectively be applied to find the signal of a family. The efficiency of finding the signal of a given family within a search can be as high as 100\% depending on slope to be searched and $a_{\rm w}$ width considered for the search. Therefore, any detection made with this technique can be also partly judged on how unlikely of a detection it was to make. Finally, as a general guide, we built the panel shown on figure \ref{fig10} bottom right. This panel shows the averaged value of the detection efficiency for all ($a_{\rm w},K$) pairs regardless family to background levels. 

Using an optimized $a_{\rm w}$ for a given slope $(K)$ does not guarantee a clear identification of the family. To really distinguish between family members and background components, as well as precisely determine the slope and age of the family, a more dedicated analysis is required. Such analysis should consider physical parameters, SFDs, may require additional HCM focused tests, etc… \citep[see][]{vokrouhlicky2006,walsh2013,dykhuis2015}. One important layer that could be added to the present work is the statistical test introduced in \citet{delbo2017} and successfully applied by \citet{delbo2019} to relate the signals obtained from their scoring maps with the ancient families Athor and Zita. However, in the present work, our main goal is to simply characterize the efficiency of the V-shape method in finding signals of families embedded in the inner main asteroid belt, i.e., candidate families that will later be confirmed (or not), and better constrained, by using additional set of data as well as other possible characterization tools. Therefore, we leave this task of including the statistical test for a follow up of this work. 

In other words, once we define the optimal range of $a_{\rm w}$ values that would most efficiently return signal of families of a given age, in a follow up work we would be able to apply for the optimal pairs ($a_{\rm w},K$) the statistical test, thus, eventually assessing the reliability of the method. Then, once this is done (in a similar way for the $dC$-$method$ and for both- and one-side methods), we would be able to deploy this tools to scan the entire inner Main Belt, searching for signals of all known and missing families. Also, by characterizing the expected detection efficiency for each set of family age and density, this method should be able to tell us how many of such families are {\it not} being detected with this technique. As an example, imagine this method returns with the signal for a family of slope 2 ${\rm km^{-1}au^{-1}}$ in the range of D.E. = 0.7 for a given signal-to-noise. This would mean that about 30\% of the $K =$ 2 ${\rm km^{-1}au^{-1}}$ family inventory found are still hidden and is possible to be detected. Still, we once again stress that this will only be possible after we assess the reliability of the method, which we leave for future work.

Another important point to note from the values of $a_{\rm w}$ is that for very old families, $a_{\rm w}$ may be as wide as 0.5 au. This means that the width of the $a_{\rm w}$ sliver for finding very ancient families is optimal when $a_{\rm w}$ equals almost the entire inner main belt. This makes sense if we consider that by grabbing as much objects of the inner main belt within $[ K|a-ac|$ and $K(|a-ac|+a_{\rm w}) ] = N_{in}$ as we can (objects above the nominal V) and dividing this number by the number of objects below the nominal V, $N_{out} = [ K|a-ac|$ and $K(|a-ac|-a_{\rm w}) ]$, the contrast $N_{in}^2/N_{out}$ returned by our method will be maximized. Of course, depending on the density of the background and how isolated the family is in the data set considered, the edge of the family can be so clear that even a narrower sliver can detect its signal. A good example of that is the case for the ancient families found by \citet{delbo2017,delbo2019}, where the selection of the data set used allowed the families found to be recognized even by eye. This can also, of course, be problematic when the search becomes so wide that it encounters the edges of a region of the asteroid belt, thus adding to the complications of finding very old families.

An approach that would also possibly result differently for the $a_{\rm w}$ prediction is weighting the objects diameters within the data set used for the search. This technique, not considered in the present work, was developed by \citet{bolin2018a} and consists in giving more importance for the large diameter objects above the nominal V, rather than accounting for every object as we do. Therefore, although the strategy used by \citet{delbo2017,delbo2019} is correct and should be used in every case where physical properties are well known, as well as that from \citet{bolin2018a} (certainly a much better approach to characterize the precise slope of the family/signal found by any method or data selection), our method intention is to characterize the best set of parameter that most efficiently return a signal that well represents a family slope and center, while being more general and less dependent on data selection. 

As a final note, recall that what is shown in figure \ref{fig10} is the optimal value for detection with efficiency larger than 50\% for a both side scoring approach \citep[for example considered a left side scoring approach; see section \ref{awKlr}]{delbo2017}. Therefore, this does not imply that one would never be able to find the signal of an ancient family with a narrower $a_{\rm w}$ width in a both side scoring approach. What our method implies is that the likelihood of finding such an ancient family adopting a narrower $a_{\rm w}$ width in a both side scoring is less than 50\%, and so, not considered by our criteria.

\subsection{$a_{\rm w}$ as function of K; right/left side scoring in the $a_{\rm w}$-$method$}\label{awKlr}

 \begin{figure*}
 	\centering
 		\includegraphics[scale=0.35]{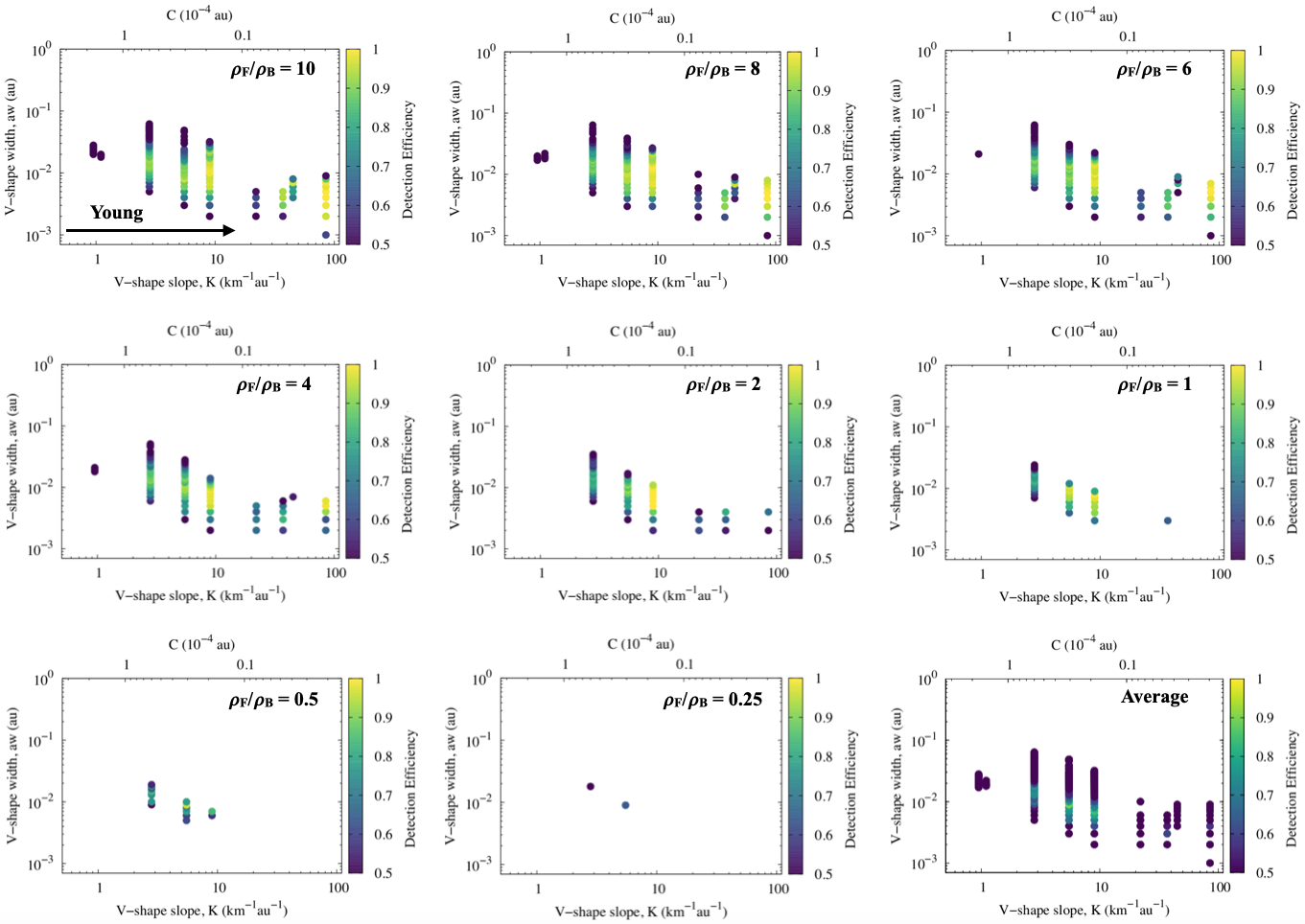}
 	\caption{Detection efficiency for $a_{\rm w}$ as a function of $K$ within $a_{\rm w}$-$method$ one side scoring.}
 	\label{fig11}
 \end{figure*}

When only one side of a V-shape is used for scoring a similar pattern of optimal widths are found, but with slightly depressed efficiencies. Similarly to figure \ref{fig10}, figure \ref{fig11} shows how the width $a_{\rm w}$ relates to the slope $K$ of a family within the $a_{\rm w}$-$method$, but now when considering only a right side scoring (recall our synthetic families evolve into a right side family at old ages, section \ref{famevodin}). Once again, in figure \ref{fig11} we plot the detection efficiency for all pairs ($a_{\rm w},K$) that returned D.E. $\geq$ 0.5. The same similarity observed in figure \ref{fig10} can be seen in figure \ref{fig11}, that is, despite some small differences within the panels in figure \ref{fig11}, the shape of the curves that relate $a_{\rm w}$ with $K$ and the range of $a_{\rm w}$ widths that return D.E. $\geq$ 0.5 for a given slope $K$ are very similar in all panels. However, for the cases presented in figure \ref{fig11}, detection becomes less and less likely for $\rho_F/\rho_B <$ 2 ($\rho_F/\rho_B <$ 4 for old families).

As before, due to the absence of randomness in scattering of points in the panels of figure \ref{fig11} we can conclude that regardless of size or density that a given family may have respect to a certain background, there is an optimal range of $a_{\rm w}$ widths that better characterize the method (making it over 50\% effective in the application to find the signal of a family in the $a_{\rm w}$-$method$ when a left/right side scoring is performed), and so, as a general guide, we built the panel shown in figure \ref{fig11} bottom right. This panel shows the averaged value of the detection efficiency for all ($a_{\rm w},K$) pairs regardless family to background levels.

Different from the both side scoring approach, in the case of one side scoring approach (right side in this case) the optimal value of $a_{\rm w}$ seems less sensitive for the value of $K$. For ancient families $K \leq$ 1 ${\rm km^{-1}au^{-1}}$ the optimal value of $a_{\rm w}$ is of the order of $10^{-2}$, raging around $a_{\rm w} =$ 0.02-0.03 au (figure \ref{fig11}). Remarkably, $a_{\rm w} =$ 0.03 au was the precise value adopted by \citet{delbo2017} to find a $K \approx$ 0.6 ${\rm km^{-1}au^{-1}}$ primordial family when considering a left side scoring approach (see \citealp{delbo2017} supplementary material for details). Additionally, \citet{delbo2017} pointed out in their supplementary material that values of $a_{\rm w}$ in the range of 0.01-0.05 au all gave similar results. These similarities cannot be just serendipitous and thus we argue that our results help to explain the \citet{delbo2017} primordial family detection, and the \citet{delbo2017} result supports this work. 

\subsection{$dC(dK)$ as function of C; both- and one- (right/left) side scoring in the $dC(dK)$-$method$}\label{CdC}

 \begin{figure*}
 	\centering
 		\includegraphics[scale=0.35]{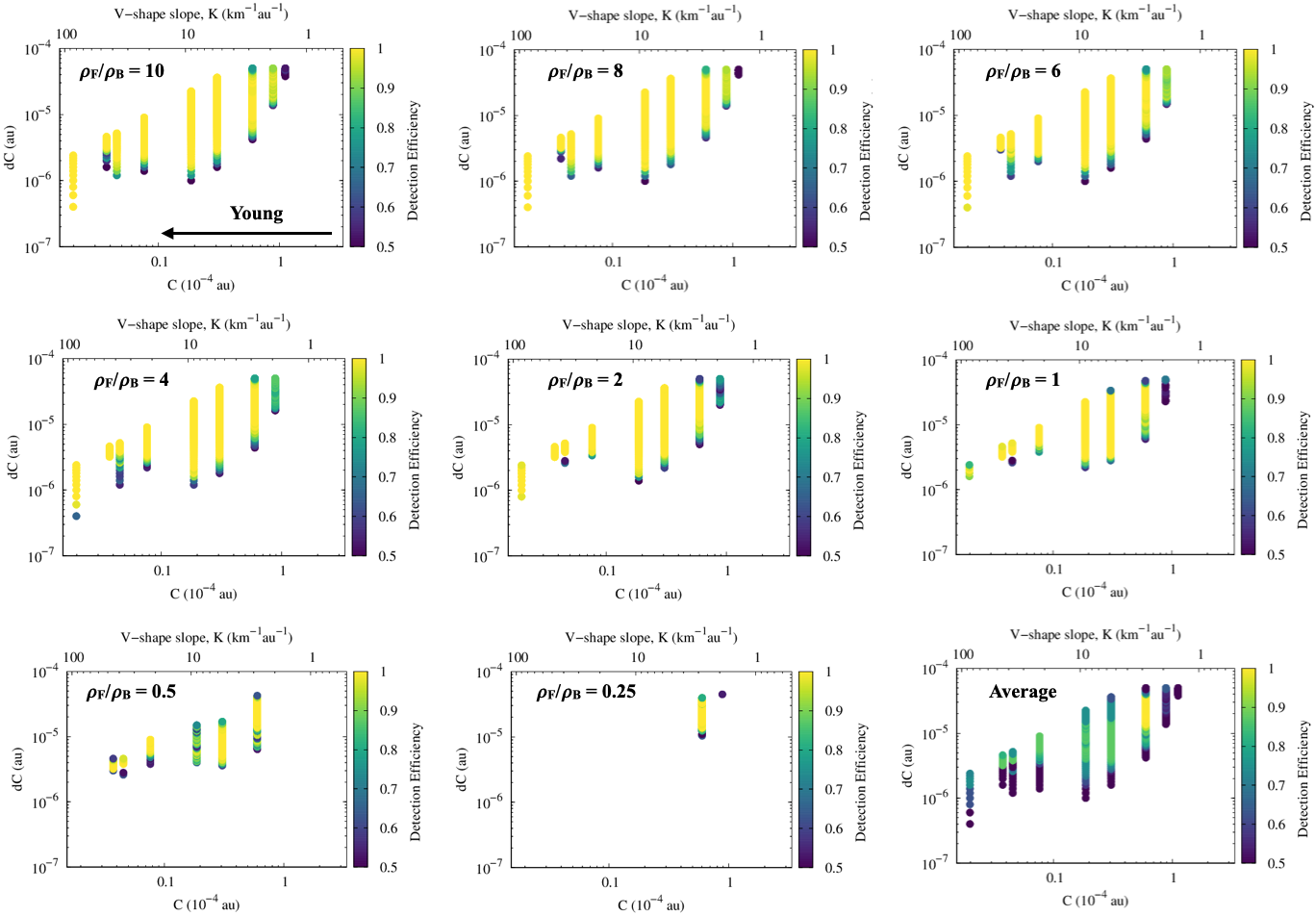}
 	\caption{Detection efficiency for $dC$ as a function of $C$ within $dC$-$method$ both side scoring.}
 	\label{fig12}
 \end{figure*}

 \begin{figure*}
 	\centering
 		\includegraphics[scale=0.35]{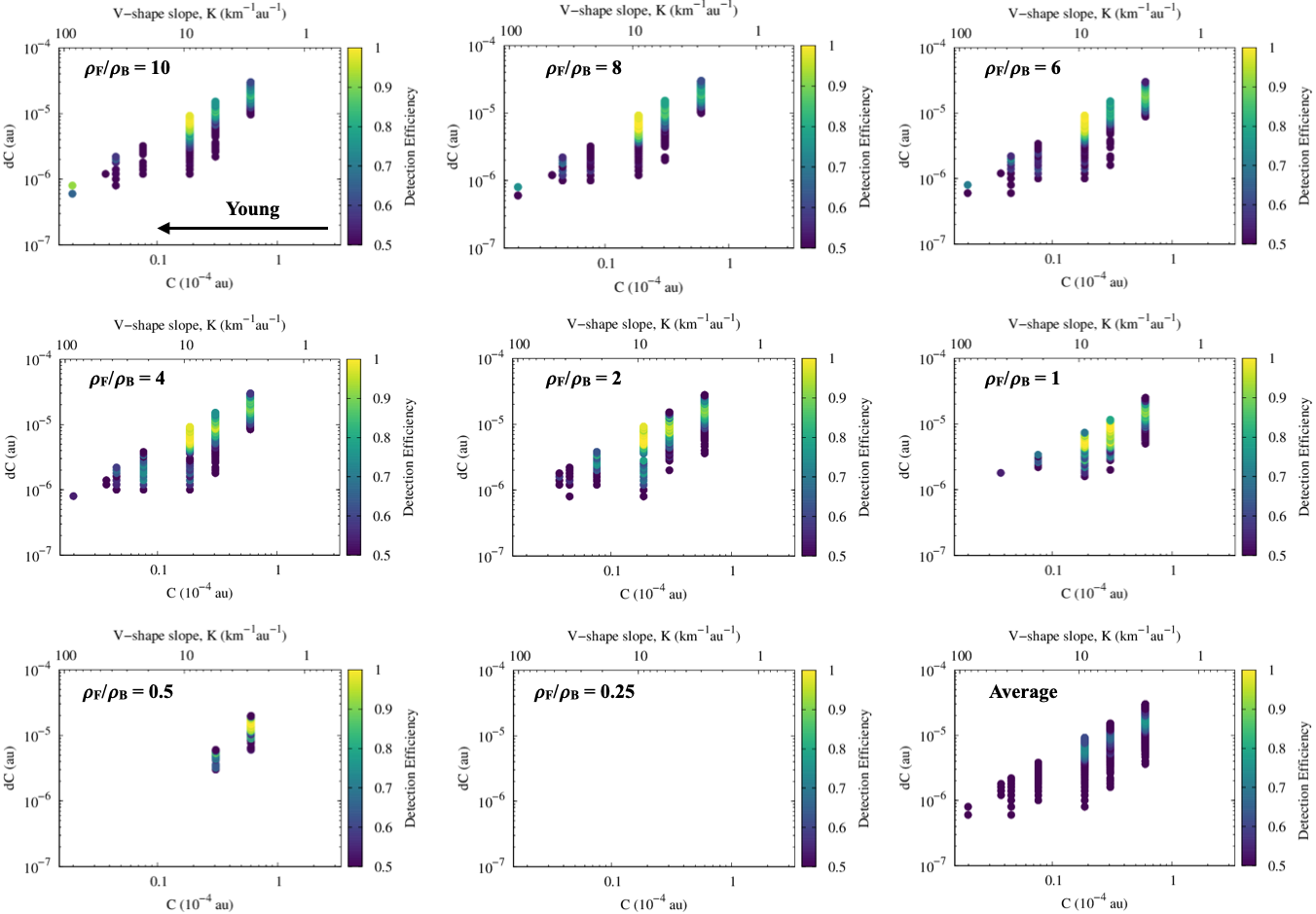}
 	\caption{Detection efficiency for $dC$ as a function of $C$ within $dC$-$method$ one side scoring.}
 	\label{fig13}
 \end{figure*}

Now we turn our attention to the $dC(dK)$-$method$ (recall that as $K$ and $C$ are a function of each other, both $dC$ or $dK$ methods are equivalent and based in small changes in the nominal slope). The $dC$-$method$ was pioneered by \citet{bolin2017,bolin2018b,bolin2018a} and makes direct use of the characteristic value $C$ \citep{vokrouhlicky2006,walsh2013} while the $dK$-$method$ was applied by \citet{delbo2019}. Here we will focus on the $dC$-$method$ \citep{bolin2017,bolin2018b,bolin2018a}, while a translation for the $dK$-$method$ should be straightforward.

Figures \ref{fig12} and \ref{fig13} are similar to figures \ref{fig10} and \ref{fig11} respectively. However, here we plot how the sliver $dC$ relates to the characteristic value $C$ of a family within the $dC$-$method$ (the nominal slope $K$ is also shown in the top axis of each panel for reference). This is done when considering a both side scoring (figure \ref{fig12}) and only a right side scoring (figure \ref{fig13}). As before, in both figures \ref{fig12} and \ref{fig13} we plot the detection efficiency for all pairs ($dC,C$) that returned D.E. $\geq$ 0.5. As can be seen from figures \ref{fig12} and \ref{fig13}, once again, despite some small differences within the panels with different values of signal-to-noise, the shape of the curves that relate $dC$ with $C$ and the range of $dC$ slivers that return D.E. $\geq$ 0.5 for a given characteristic value $C$ are very similar in all panels.

Despite the fact that detection in these cases becomes less and less likely for $\rho_F/\rho_B <$ 1 for both scoring methods, we once again notice a remarkable absence of randomness in scattering of points in the panels of figures \ref{fig12} and \ref{fig13}. Therefore, we can also conclude that there is an optimal range of $dC$ slivers that will be best suited for the $dC$-$method$, as we found for the $a_{\rm w}$-$method$, whenever performing a both- or a left/right-side scoring. The fact that an optimal value of $dC$ exists for this method is not totally new as it was first pointed out by \citet{bolin2017}. However, here we are quantifying this relation ($dC,C$) so as we are able to fully characterize the $dC$-$method$ as we called. Thus, as it has been made in figures \ref{fig10} and \ref{fig11}, as a general guide, we incorporated to figures \ref{fig12} and \ref{fig13} a panel (bottom right in these figures) that shows the averaged value of the detection efficiency for all ($dC,C$) pairs regardless family to background levels. 

The analyses and conclusions from figures \ref{fig12} and \ref{fig13} are very similar to the discussion in the previous two sections. That is, younger families (steeper slopes) require smaller $dC$ and older families (shallower slopes) larger $dC$. Also, the likelihood of detection is higher for both side scoring when compared to left/right scoring.

One big difference that exists between $a_{\rm w}$-$method$ and $dC$-$method$ is that while in the $a_{\rm w}$-$method$ the detection efficiency seems to be indifferent to the slope that is being scored, in the $dC$-$method$ this is not true. In no case that we performed a $dC$-$method$ did we score $K \leq$ 1 ${\rm km^{-1}au^{-1}}$ with at least 50\% efficiency. The question that then raises is how was \citet{delbo2019} able to find two ancient families $K \approx$ 1.72 ${\rm km^{-1}au^{-1}}$ (Athor) and $K \approx$ 1 ${\rm km^{-1}au^{-1}}$ (Zita) with the $dK$-$method$? Do our results rule out the findings by \citet{delbo2019}? The answer for the latter question is no and the reason relies on the answer for the first question. According to our figures \ref{fig12} and \ref{fig13}, $K \approx$ 1.72 ${\rm km^{-1}au^{-1}}$ (Athor) is well within the range of slopes that can be identified by the $dC(dK)-method$ with more than 50\% efficiency. As for the case of $K \approx$ 1 ${\rm km^{-1}au^{-1}}$ (Zita), although we did not get any D.E. $\geq$ 50\%, this does not mean D.E. = 0. It only means detection is very unlikely. Also, as previously discussed in section \ref{awK}, by using data from physical parameters, \citet{delbo2019} was able to reduce their sample of asteroids in such a way that one could almost detect the family by eye (see figures 5 bottom panel and A.1 in their work). Again, as also already discussed in section \ref{awK}, this is a valid approach and should be used whenever possible.

Before we draw our final conclusions, we dedicate the next section to run our optimal $a_{\rm w}$ and $dC$ values on the same data \citet{delbo2017,delbo2019} considered and so, see if we are able to find with our characterized methods the same ancient/primordial families \citet{delbo2017,delbo2019} have found, as well as to see if we can improve their detection signals.

\section{Practical application: Testing results by \citet{delbo2017,delbo2019}}\label{application}

 \begin{figure*}
 	\centering
 		\includegraphics[scale=0.4]{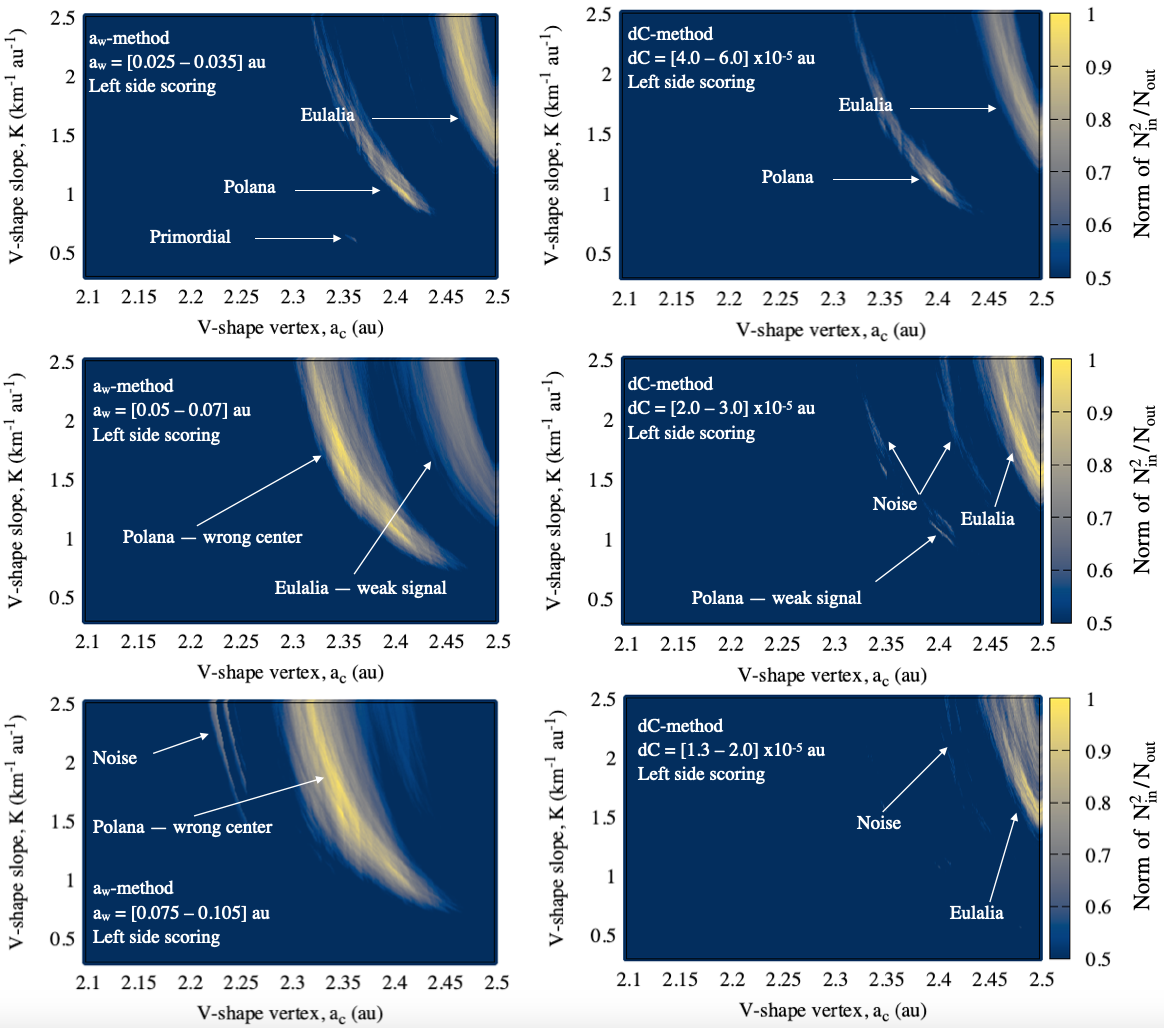}
 	\caption{Score map for different values of $a_{\rm w}$ in the $a_{\rm w}$-$method$ (left) and $dC$ in the $dC$-$method$ (right), considering a left side scoring in the data from \citet{delbo2017}. Top: applying our optimal values for $a_{\rm w}$ and $dC$. Center: using the range of $a_{\rm w}$ times 2 and the range of $dC$ times 0.5 (from the top panels). Bottom: using the range of $a_{\rm w}$ times 3 and the range of $dC$ times 0.33 (from the top panels).}
 	\label{fig14}
 \end{figure*}

 \begin{figure*}
 	\centering
 		\includegraphics[scale=0.35]{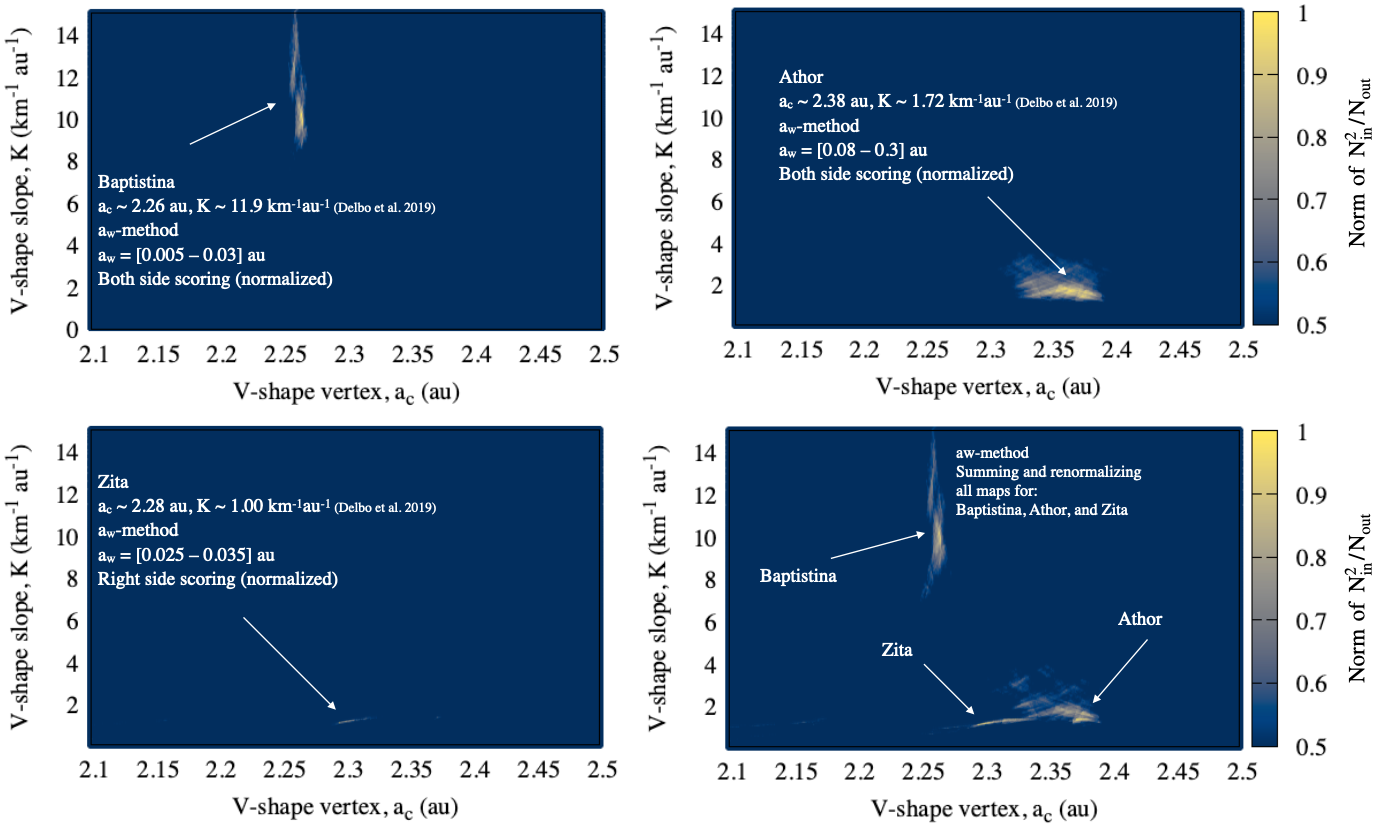}
 	\caption{Score map applying our optional interval of $a_{\rm w}$ in the $a_{\rm w}$-$method$ to the data from \citet{delbo2019}. Top left: Baptistina family found. Top right: Athor family found. Bottom left: Zita family weakly found. Bottom right: the signal for all three families in the same ($K,a_c$) space with minimum noise.}
 	\label{fig15}
 \end{figure*}

 \begin{figure*}
 	\centering
 		\includegraphics[scale=0.35]{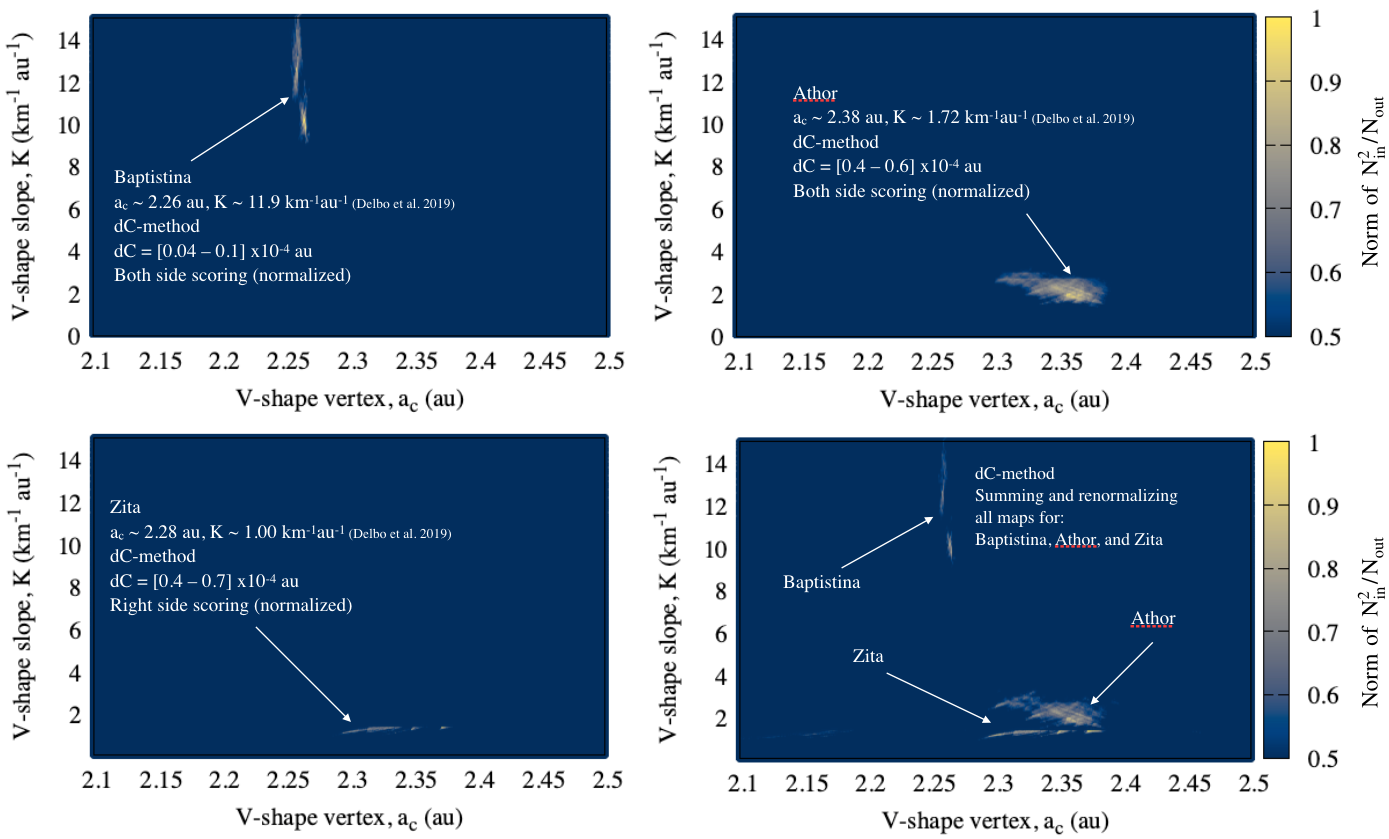}
 	\caption{Score map applying our optional interval of $dC$ in the $dC$-$method$ to the data from \citet{delbo2019}. Top left: Baptistina family found. Top right: Athor family found. Bottom left: Zita family weakly found. Bottom right: the signal for all three families in the same ($K,a_c$) space with minimum noise.}
 	\label{fig16}
 \end{figure*}

Once we now have defined the optimal values for $a_{\rm w}$ and $dC$ in both $a_{\rm w}$-$method$ and $dC$-$method$ from our synthetic families and background, let us apply these values for a real main belt data set. For this task we will use the same data considered by \citet{delbo2017,delbo2019}. The reason for this choice relies on the fact that \citet{delbo2017,delbo2019} found one or two very old families embedded in a region with families of different ages, i.e. \citet{delbo2017} found a primordial family ($K \approx$ 0.6 ${\rm km^{-1}au^{-1}}$) along with detecting Polana ($K \approx$ 1.1 ${\rm km^{-1}au^{-1}}$) and Eulalia ($K \approx$ 1.7 ${\rm km^{-1}au^{-1}}$) and \citet{delbo2019} found two ancient families, Athor ($K \approx$ 1.71 ${\rm km^{-1}au^{-1}}$) and Zita ($K \approx$ 1 ${\rm km^{-1}au^{-1}}$) while also detecting the young Bapstina ($K \approx$ 11.9 ${\rm km^{-1}au^{-1}}$) family. As we can see from the values of $K$ within all previous families, in \citet{delbo2017} all families are of similar slopes and so, Fig. 2 in \citet{delbo2017} is clear, i.e. with good detection signals. 

However, the same is not true for the families presented in \citet{delbo2019}. Athor and Zita are of similar slopes, but Bapstina has a much higher slope than them. As a result of such difference in slopes, Fig. 3 in \citet{delbo2019} is very fuzzy, i.e., although the signal for Athor family is very clear, the same is not true for Baptistina and Zita. This, on the other hand, is expected from our present work due to two reasons. First, once we showed that $a_{\rm w}$ (or $dC$) is a function of $K$ (or $C$) to find families of different slopes one would need different values of $a_{\rm w}$ or $dC$ \citep[not considered by][]{delbo2019}. Second, recall that \citet{delbo2019} considered the $dK$-$method$ and that Zita is a right side family  \citep[see figures Fig. 5 bottom and A1 in][]{delbo2019}. This means that $dC$ for such $K \approx$ 1 ${\rm km^{-1}au^{-1}}$ is out of our 50\% detection efficiency range (figure \ref{fig13}), thus such a signal, if correct, is expected to be weak. Therefore, in the following we will apply our optimal values for $a_{\rm w}$ and $dC$ to \citet{delbo2017,delbo2019} data set and see how the detection can be improved. One additional positive point in applying our optimal values to \citet{delbo2017,delbo2019} data set is that, because we know where the real families are inside such a data set, we can not only test \citet{delbo2017,delbo2019} results, but also test our results and better characterize what, within a search in a real case, would be detection signal or noise.

In figure \ref{fig14} we show a proof-of-concept for our method when applied to \citet{delbo2017}. In the left panels of figure \ref{fig14} we have applied the $a_{\rm w}$-$method$ and on the right panels the $dC$-$method$. One thing not discussed so far, but used to create figure \ref{fig14} and all the subsequent figures \ref{fig15} and \ref{fig16}, is that, in order to decrease the sensitivity of the results due to changes in $a_{\rm w}$ or $dC$ (figure \ref{fig9}) we did not perform the scoring in only one value of $a_{\rm w}$ or $dC$. Instead, we considered a range of $a_{\rm w}$ and $dC$ values within the optimal intervals shown in figures \ref{fig10}, \ref{fig11}, \ref{fig12}, and \ref{fig13}. The precise way we did is: $i$) within the desired range of slopes to search for, we chose the minimum and maximum optimal values of $a_{\rm w}$ or $dC$ from that interval; $ii$) we defined a $\Delta a_{\rm w}$ and $\Delta dC$ such we would consider $N$ values of $a_{\rm w}$ and $dC$ within minimum and maximum values; $iii$) for each $a_{\rm w}$ or $dC$ we performed a scoring (left side in the case of figure \ref{fig14}), built a score map, and normalized such score map; $iv$) then, we summed all $N$ normalised scoring maps and re-normalized the addition. This was done to maximize the best scoring region, i.e., imagine all $N$ searches give similar but different normalized score maps. By adding one to another the similar normalized signals would become larger than unity (approaching N), whereas possible noise or nearby regions would add up to values smaller than unity (or at least much smaller than N). Thus, by re-normalizing the new summed scores all the noise around the main signal is strongly decreased. This does not imply that we are able to decrease noise from the data set, but from the searching method itself. Therefore, by doing this, we decrease the sensitivity of our results due to the specific choices of $a_{\rm w}$ or $dC$ (choices that according to our study could be bad, leading to wrong or fuzzy signals, or where efficiency is much smaller than 50\%). Finally, we also show in the following figures only normalized scores above 0.5 so as to decrease the level of noise and only rely on signals that are strong enough to be higher than 0.5.

The top panels of figure \ref{fig14} shows the case where the optimal values of $a_{\rm w}$ and $dC$ range from 0.025 au to 0.035 au and 4$\times$10$^{-5}$ au to 6$\times$10$^{-5}$ au respectively. These are the optimal range of $a_{\rm w}$ and $dC$ values from figures \ref{fig11} and \ref{fig13} when performing a one side (left, \citealp{delbo2017} only found left side families) search for the slopes considered. As one can see, the crispness of the results presented in our figure \ref{fig14} is a great improvement of what is shown by Fig. 2 in \citet{delbo2017}. Additionally, as previously discussed, the value of $a_{\rm w}$ used by \citet{delbo2017} is within the range of optimal $a_{\rm w}$ found in this work. Thus, it is not a surprise that we have found the same families with the $a_{\rm w}$-$method$. It is also not a surprise that the primordial family found by this method has a very weak signal when compared to the other two (D.E $\approx$ 50-60\% for $K \approx$ 0.6 ${\rm km^{-1}au^{-1}}$). With the $dC$-$method$ however, we were not able to find the primordial family at $a_c \approx$ 2.36 au and $K \approx$ 0.6 ${\rm km^{-1}au^{-1}}$ with a scoring $>$ 0.5. This is also expected due to the fact that there is no optimal $dC$ defined for $K \approx$ 0.6 ${\rm km^{-1}au^{-1}}$ with efficiency larger than 50\% in our work (see however figure \ref{fig16}, case for Zita family). Note though, that the fact that there is no signal for the primordial family in the $dC$-$method$ as we presented does not imply that the signal does not exist at all (it may appear for scoring levels below 0.5, which are not shown here for consistency). 

In order to determine how precise or sensitive our optimal values are, we also show in figure \ref{fig14} center and bottom panels, the corresponding scoring maps for the cases where we considered the range of $a_{\rm w}$ 2 times the optimal interval and 0.5 times the optimal interval for $dC$ (center panels), as well as 3 times the optimal interval for $a_{\rm w}$ and 0.33 times the optimal interval for $dC$ (bottom panels). It is clear from figure \ref{fig14} center and bottom panels that being outside the optimal interval of $a_{\rm w}$ or $dC$ even by a little may disturb the score map leading to misplacement of signals, thus generating noise.

Noise as we call in figure \ref{fig14} (as well as in the following figures \ref{fig15} and \ref{fig16}), is a clear result of {\it edge effects}. Therefore, we can conclude from these figures that the correct choices of $a_{\rm w}$ and $dC$ ($dK$) (which we are characterizing in the present work) also provides a way to avoid, or at the very least, decrease the influence of {\it edge effects}.

Continuing with our real case tests, we now turn our attention to the findings presented by \citet{delbo2019}. Figures \ref{fig15} and \ref{fig16} are another proof-of-concept showing how the appropriate choice of the optimal intervals for $a_{\rm w}$ (figure \ref{fig15}) or $dC$ (figure \ref{fig16}) can be used to find families of different slopes with minimum noise. Top left panels of figures \ref{fig15} and \ref{fig16} show respectively the optimal interval of $a_{\rm w}$ and $dC$ used to find Bapstina family (Athor and Zita as well as background asteroids were all embedded in the data set). Similarly, figures \ref{fig15} and \ref{fig16} top right panels show the signal found for Athor family without noise or pollution from Baptistina family. The reason behind why both left and right panels in the top of figures \ref{fig15} and \ref{fig16} present very clear signals for two very different slopes is a proof of our concept that specific slopes within a single data set can be found by the correct choice of $a_{\rm w}$ or $dC$. In this case, the signal of Athor family is particularly invisible for Baptistina scoring and vice-versa.

Although both $a_{\rm w}$-$method$ and $dC$-$method$ were able to find Bapstina and Athor very clearly in a both side scoring, Zita was not found in any attempt. The reason for that is because Zita (a right side ancient family) has a very weak signal when compared to Baptistina and Athor. Therefore, in order to be able to detect Zita signal we had to perform a right side scoring considering only slopes up to $K \approx$ 1.5 ${\rm km^{-1}au^{-1}}$, instead of the whole interval from 0.1 ${\rm km^{-1}au^{-1}}$ to 15 ${\rm km^{-1}au^{-1}}$ (bottom left panels of figures \ref{fig15} and \ref{fig16}; although we show the panels for Zita up to $K =$ 15 ${\rm km^{-1}au^{-1}}$, we filled the grid $K,a_c$ with zeros for $K >$ 1.5 ${\rm km^{-1}au^{-1}}$). Due to the fact that the optimal fit for $a_{\rm w}$ in this case is similar to Baptistina, a search within the entire interval of slopes would certainly return a higher signal for the right side of the Bapstinia family. Additionally, for intermediate intervals of slope, too much noise from Athor would be present in the scoring map. Similarly, in the $dC$-$method$, the optimal value for $dC \approx$ 4$\times$10$^{-5}$ au to 7$\times$10$^{-5}$ au (inferred from figure \ref{fig13} once there is no real optimal $dC$ value for $K <$ 1 ${\rm km^{-1}au^{-1}}$ for a one side $dC$ scoring) is equal the optimal $dC$ range for Athor. Thus, searching for slopes higher than $K \approx$ 1.5 ${\rm km^{-1}au^{-1}}$ would certainly find Athor instead of Zita (which even in these cases has a very weak signal).

Finally, to be consistent with the idea that by using the appropriate optimal intervals of $a_{\rm w}$ or $dC$ we could clearly find families of different ages (slopes) within a single data set, we created the bottom right panel in figures \ref{fig15} and \ref{fig16}. For this panel we summed all three normalized panels for Baptistina, Athor and Zita and then once again re-normalized the resulting score map. By doing so, we can clearly see the signals for Bapstina and Athor (very different slopes) as well as the signature of Zita's signal within one single plot.

Last but not least, our efforts show a slight preference for old families to be found by the $a_{\rm w}$-$method$ (figures \ref{fig10}, \ref{fig11}, \ref{fig12}, \ref{fig13}, and \ref{fig14} top panels). Other than that, both $a_{\rm w}$-$method$ and $dC$-$method$ are equally powerful in finding signals of different slope families, since considering the right interval of optimal $a_{\rm w}$ and $dC$ values presented in our work.

\section{Conclusions}\label{conclusion}

This work provided a rigorous inspection of the behavior of different types of V-shape asteroid family finding methods over a wide range of parameters and asteroid family properties. The V-shape techniques are found, as expected to be very sensitive to the search parameters that vary with the age, or V-shape slope, of the targeted family. These dependencies were quantified and provide a guide for future users to optimize their searches.

The ability of the techniques also depends strongly on the asteroid family itself; both its V-shape slope and also its relative density of objects compared to the background asteroid belt. This is quantified and should serve as a guide for future users to understand how likely or unlikely a detection should be for a specific family. Moreover, by performing additional experiments, not shown, where we steepened the slope of our background population's SFD, we found that different choices of SFD have a minor effect in our characterization methods. More important is indeed the contrast between family and background sizes, measured by means of their densities in this work.  

Specifically, the V-shape searching tool was applied to recently discovered inner main belt asteroid families and all were detected using the optimal search parameters for each. The next exercise would be to add statistical analysis, as well as other possible de-biasing techniques, and try to access the reliability of the method detection, rather than simply characterize its efficiency (which in turn was an essential first step towards accessing de-bias and reliability).

Once more work has been done, and de-bias and reliability could finally be accessed, we will likely be able to deploy these tools blindly over different regions of the asteroid belt in an attempt to find previously unknown families. 

\section*{Acknowledgements}
The authors are very thankful to Bojan Novakovi\'{c} and Miroslav Bro\v{z} for their very detailed and constructive reviews that greatly improved this work.
R.D. and K.W. were supported by the National Science Foundation, grant 1518127. This work used the Extreme Science and Engineering Discovery Environment (XSEDE), which is supported by National Science Foundation grant number ACI-1053575. The work of M.D. was supported by the ANR ORIGINS (ANR-18-CE31-0014). Here we made use of asteroid physical properties data from \url{https://mp3c.oca.eu/}, Observatoire de la C\^ote d'Azur.

\bibliographystyle{cas-model2-names}

\bibliography{cas-refs}





\end{document}